\documentclass[aps,amssymb,prl,twocolumn]{revtex4}
\usepackage{graphics}

\begin{document}
\bibliographystyle{prsty}

\title{Boundary lubrication with a glassy interface}

\author{Ana\"el Lema\^{\i}tre$^{(1,2)}$}
\author{Jean Carlson$^{(1)}$}
\affiliation{
$^{(1)}$ Department of physics, University of California, Santa Barbara, California 93106, U.S.A.}
\affiliation{
$^{(2)}$ L.M.D.H. - Universite Paris VI, UMR 7603, 4 place Jussieu - case 86, 75005  Paris - France}
\date{\today}

\begin{abstract}
Recently introduced constitutive equations for the rheology of dense, disordered
materials are investigated in the context of stick-slip experiments in boundary
lubrication. The model is based on a generalization of the shear transformation
zone (STZ) theory, in which plastic deformation is represented by a population of 
mesoscopic regions which may undergo non affine deformations in response to stress.
The generalization we study phenomenologically incorporates the
effects of aging and glassy relaxation.
Under experimental conditions associated with typical transitions from stick-slip 
to steady sliding and stop start tests, these effects can be 
dominant, although  the full STZ description is 
necessary to account for more complex, chaotic transitions.
\end{abstract}
\maketitle

\section{I. Introduction}

Advances in developing  nanometer scale technologies and devices  are intrinsically
coupled to fundamental progress in scientific  understanding of
properties of materials under atomic scale
confinement~\cite{horn81,christenson87,israelachvili88,schoen89,gee90,thompson90a,thompson90b,grannick91,ribarsky92,thompson92,bitsanis93,persson93a,stevens93,yoshizawa93a,yoshizawa93b,reiter94,persson94,melnichenko95,urbakh95,carlson96,gao96,demirel96a,gao96,gao97,luengo97,rozman98,batista98,persson00,drummond00,berthier01,drummond01,berthier02a,gourdon03,robbins03}.
On one hand, friction, fracture, and plastic deformation at microscopic scales 
influences the operation and performance of engineered systems. On the
other hand, investigation of clean, well
characterized nanoscale systems  is leading to new insights into these
underlying physical phenomena which  govern
systems under stress.

Increases in scientific computing capacity along with the development of new
experimental techniques, has recently created opportunities for progress
in both theory and measurements.
The surface force apparatus (SFA) was originally
designed to study solvation forces induced by a liquid confined between parallel
surfaces~\cite{horn81,christenson87}. Subsequently it was adapted to measure
shear forces~\cite{israelachvili88}. The SFA allows precise measurements 
on a microscopically thin layer of lubricant, separating atomically smooth
(typically mica) surfaces. Friction and/or adhesion associated
with a single asperity contact can be precisely measured.  We focus here on friction
and the associated stick-slip  instabilities.

In many practical instances friction
involves rough materials. 
However, for rough surfaces  it can be difficult 
to identify underlying mechanisms associated with
complex phenomena (e.g. irregular dynamics, and 
bifurcations). For rough materials individual asperity dynamics are {\it a priori}
combined with any
collective phenomena which may be associated with 
the population of contacts,
and the population of contacts itself is necessarily time dependent.
Rough surface measurements are thus naturally complementary to 
investigations of plasticity and rheology of isolated, 
individual asperities.
In addition, for single asperities, effects associated with
interfacial materials (e.g. lubricants) are more easily isolated.
Furthermore,
there is growing evidence that 
the frictional properties of rough surfaces even at macroscopic scales
is controlled by the plastic deformation 
of individual contacts. Indeed, experiments  have recently 
been designed which
isolate and measure the dynamics of individual asperities at a 
rough dry  interface, subject to  shear~\cite{bureau02}.

In this paper we focus on dynamics of an individual lubricated
asperity contact. We model the lubricant using a set of
constitutive equations which generalize the shear 
transformation zone (STZ) theory for
amorphous, glassy materials. In addition to the STZ equations,
the model incorporates the 
effects of glassy relaxation
{\it via} the introduction of a state variable
related to the internal free-volume (the additional state
variable may alternatively
be thought of as an out-of-equilibrium effective temperature).
The coupled STZ and free-volume 
dynamics 
were introduced previously, 
and shown to capture a range of experimental phenomena in
glassy and granular materials~\cite{lemaitre02b,lemaitre02c}.
Here the constitutive equations model
internal states of the lubricant. 
We perform a series of analytical and numerical calculations which
mimic typical SFA  experiments~\cite{yoshizawa93b,drummond00,drummond01}, 
and investigate the stationary states, 
the bifurcation diagram of the transition between stick-slip and steady sliding,
the nature of this transition (i.e.~super- or sub-critical), the emergence of chaos,
and aging of the yield stress in stop-start tests.
At this stage, we primarily map out qualitative
behaviors of the model. The compelling correspondence to existing experiments 
sets the stage for future, more detailed, quantitative 
comparison with data.

The remainder of this paper is organized as follows. 
Section II  provides  a brief overview of friction and
boundary lubrication.
Section III describes free-volume constitutive
equations and summarizes elements of STZ theory.
Section IV  contains the numerical and analytical results in scenarios
representative of SFA experimental studies.
Finally, we conclude in Section V with 
a discussion of our results,  comparisons with experiments,
and  directions for
continuing research.

\section{II. Background}
\label{sec:background}

In this section we provide a brief overview of boundary lubrication. Our 
emphasis is  on novel material properties and the  associated modeling challenges 
which arise for atomically thin, confined liquid films. 
Note that, even for interfaces with relatively simple 
features, we still lack precise quantitative, 
predictive models for friction. 
A more complete overview of recent results on friction and lubrication
can be found in~\cite{persson93a} and references therein.

In boundary lubrication a molecularly
thin film of material is confined between two parallel surfaces.
The relative motion of the surfaces is mediated by the
plastic deformation of the interfacial material 
lubricating the contact.
Molecularly thin films display specific properties which
differ from the viscous behavior of bulk materials.
Among the most noteworthy of these is the development of a yield
stress for thin films at sufficiently low temperatures.

The changes of material properties under confinement can 
largely be attributed to a liquid-solid 
transition~\cite{gee90,grannick91}.
In some cases, numerical and experimental evidence  suggests
layering occurs in the interfacial 
material~\cite{schoen89,gee90,thompson90a,ribarsky92,thompson92,stevens93,gao96,gao97}
and the situation is similar to a liquid-crystal transition. That is, the material orders,
and its deformation is expected to result from 
propagation of dislocations, 
or layer-over-layer sliding.
In other cases, no evidence of ordering of the interfacial material is observed, and 
the liquid-solid transition induced by confinement enters the very large class 
of structural glass transitions~\cite{thompson92,bitsanis93,demirel96a,gourdon03}. 
In this case the material remains amorphous (liquid-like), but 
displays solids-like properties.
The emergence of a yield stress is accompanied
by power-law viscosities,  as well as
signs of glassy aging~\cite{drummond00,drummond01,bureau02} and
anomalous response spectra~\cite{demirel96a,luengo97}.
For some lubricants, either a glassy or layering transition  can 
be observed, 
depending on features such as temperature~\cite{gourdon03},
holding time, surface roughness~\cite{gao00},
and commensurability of the surface and the film~\cite{gao97}.
Consequences of solid-like ordering within the film  include  development of a static
yield stress, and stick-slip instabilities~\cite{yoshizawa93a,yoshizawa93b}.

Boundary lubrication and the SFA experiments provide
special opportunities for theory, because the interface is well characterized and
precisely controlled, yet the system is large enough to display phenomena which also arise macroscopically.
To date, models have primarily emphasized effects associated with ordering and
interactions between the crystalline  substrate  and the lubricant.
These aim to describe effects of  layering and surface induced
order and involve (i) simple, reduced models of non-interacting particles
in an effective periodic potential induced by the surfaces~\cite{rozman98},
(ii) Ginzburg-Landau functionals which account for heterogeneous ordering~\cite{urbakh95},
or (iii)  motion of adsorbate layers in the periodic potential
associated with a regular surface~\cite{persson93a,persson93b,persson94}.

Alternative approaches focus on internal properties of the
lubricant, including contrasts between liquid and solid-like properties
and glassy behavior. 
Phenomenological rate and state friction laws 
have been  introduced~\cite{carlson96,batista98},
in which the friction  depends on the instantaneous slip rate and a 
state variable. The state variable models the collective dependence of friction
on the internal degrees of freedom of the lubricant.
This approach assumes the fluctuations are sufficiently self-averaging
that microscopic degrees of freedom in the boundary layer can be ignored.
This simplifying assumption was inspired by a large body of work in dry
friction where rate-and-state
formulations~\cite{dieterich78,dieterich79,ruina83,dieterich94}
have shown to be useful to account for experimental data  including
stick-slip instabilities~\cite{rice83,rice86,baumberger94,heslot94,baumberger95}.
State variables can be motivated by experimental observations or molecular
dynamics. For dry friction the state variable is 
related to the average lifetime of 
individual contacts, whereas in boundary lubrication the 
state variable is loosely connected to the degree of 
internal order in the lubricant. However, the friction
laws these underlying mechanisms inspire are based more on macroscopic, thermodynamic-like
criteria, than on the underlying microscopic physics at the interface.

Our model aims to provide a microscopically motivated description
of the macroscopic forces which arise  when the amorphous, interfacial
lubricant is subject to shear.
It relies on the assumption that the deformation of amorphous materials
is controlled primarily by excluded volume effects, 
which dominate over fine details of molecular interactions. 
This is an old idea, advocated by Struik in the 70's~\cite{struikbook76},
and supported by striking similarities between very different
amorphous systems. 
As a result, a relatively simple account of visco-plasticity is expected
to hold for wide classes of amorphous systems.
In the conclusion, we discuss how the equations we study here
may account for the behavior of not only lubricants, but also
sheared granular materials~\cite{nasuno97,nasuno98,geminard99,losert00}.

For the theorist, boundary lubrication 
has some special, simplifying features which in many
respects make it an ideal template to study
plasticity of amorphous materials.
The interfacial lubricant  layer is sufficiently thin that certain
bulk phenomena,  such as strain localization, appear to be avoided.
As a consequence boundary lubrication is, in fact, one of the few experimental setups 
where {\it homogeneous} constitutive equations can be directly tested.
In contrast, for glassy bulk materials
deformation organizes in shear bands, usually 
a few particle diameters thick, 
and the strain-rate measured in an experiment averages over a non-uniform field.
The origin of strain localization is poorly understood, 
but clearly a more complete
understanding of simple homogeneous flows must be established first.
No shear banding is expected to occur in nano-confined films
because the deformation is already confined at 
scales equal to or smaller than that which
would be expected for the shear band. 
The SFA experimental setup resembles closely the sheared strip 
used in recent numerical studies of relaxation in glasses~\cite{berthier01,berthier02a}.

There are, however, several persistent 
challenges associated with  the SFA. 
Firstly, an astonishingly wide range of phenomena 
have been observed in this system for different lubricants 
under different conditions. The range of behaviors
remains a puzzle  which is difficult to piece together 
in the absence of a systematic
theory that clearly captures the most basic  observations.
The variability of experimental observations   might
be attributed to inherent difficulties in the preparation of samples 
of any glassy material, due to the effects of aging.
Certainly capturing the full spectrum of properties
is a long term target for theoretical models.
Secondly, it is  difficult to increase the stiffness of 
the apparatus beyond $\sim 3500 N/m$~\cite{delphine}.
A salient feature of boundary lubrication, which directly results from this 
finite stiffness, is the emergence of stick-slip instabilities at 
low velocity~\cite{gee90,thompson90b,yoshizawa93a,yoshizawa93b,demirel96b,drummond00,drummond01,gourdon03}.
In the most dramatic cases, the transition to stick-slip 
is accompanied by chaotic behavior~\cite{drummond01}.
The main consequence of the stick-slip instability 
(and of the finite scanning length of the experimental apparatus), 
is that the SFA can provide stationary data only for limited ranges of strain-rates.
Thus  for a wide range of parameters,
information about the interfacial material can be gathered only
through the observation of instabilities and transient dynamics.
Therefore, it is essential to treat this finite stiffness explicitly in any theoretical 
approach. 
Ultimately, rather than hampering our understanding of structural glasses, 
stick-slip instabilities and transient dynamics provide us 
much richer data than simple stationary states.
These instabilities may help clarify important, general issues associated with
the out-of-equilibrium properties of structural glasses.

The model we study in this paper is 
an intermediate statistical theory 
recently proposed~\cite{lemaitre02b,lemaitre02c,lemaitre03}
for sheared structural glasses.
Here ``intermediate'' refers to the fact our model falls
between the thermodynamic-like, 
phenomenological rate and state descriptions which have
been proposed for friction at dry and lubricated interfaces, and
an atomistic statistical mechanical description of the
lubricant, which takes into account detailed microscopic,
molecular interactions.
The model begins with  shear transformation zone (STZ) theory~\cite{falk98,falk00},
which was inspired by molecular dynamics simulations of
fracture in amorphous materials, and
which provides a microstructural description
of shear-induced rearrangements~\cite{falk98,falk99,falk00}.
STZ theory accounts for the emergence of a yield 
stress in amorphous materials
through introduction of state variables 
characterizing the anisotropy of structural arrangements. 
Drawing on earlier approaches to describing  creep in metallic
alloys~\cite{spaepen77,argon79a,argon79b,spaepen81,argon83},
STZ theory models local shear rearrangements as activated processes, 
controlled by local density fluctuations. 
This leads to the introduction of free-volume activation factors,
in the spirit of early theories of the glass transition~\cite{doolittle51,doolittle57,turnbull61,turnbull70,grest81}.

In~\cite{lemaitre02b} and subsequent works~\cite{lemaitre02c},
it was noted that previous approaches treat free-volume as a fixed
parameter, although it clearly varies as the material dilates or contracts.
In granular materials, for example, density relaxations have been observed
and precisely characterized experimentally~\cite{nowak98},
inspiring several models of slow relaxations~\cite{boutreux97,nowak98,stinchcombe02}.
Dilatancy in granular material is also
involved in the definition of frictional properties 
and stick-slip transitions~\cite{thompson91}.
Slow relaxation of volume or enthalpy are ubiquitous 
features accompanying the glass transition~\cite{struikbook76}.
We expect more explicit  modeling of
the time-dependence of density fluctuations 
may capture the emergence of a wide range of glassy properties 
near a jamming transition.
The constitutive equations introduced in \cite{lemaitre02b} were
thus developed to address the question:
How much  glassy phenomenology can be captured 
by the simplest account of free-volume dynamics, 
coupled to the dynamics of shear transformation zones?

Previously it was shown that free-volume dynamics suffices to characterize 
aging and power law rheologies, while the dynamics of shear transformation
zones are required to account for the emergence of a yield stress~\cite{lemaitre02c}.
In transient regimes, both processes may interact and contribute 
to rheological properties.
Furthermore, chaotic behavior has been observed in the SFA close to the 
stick-slip instability~\cite{drummond01}, and
it is known in over-damped frictional equations, 
that two or more state variables are necessary to understand the occurrence of chaos. 
Here, 
we did not invoke an {\it ad hoc\/} theory with multiple state variables \cite{becker00},
by were led  to it by the underlying physical mechanisms already associated with other,
relevant experimental observations.
Moreover, as chaos is  difficult to characterize experimentally, significant
insight is provided when 
observations are supplemented with theoretical models, to guide measurements and analysis.

Finally, we note that it is not {\it a priori} necessary
to identify the additional state variable as a free-volume.
It was previously noted by Falk and Langer that free-volume 
was related to the notion of Edwards' 
temperature~\cite{falk98,mehta89,edwards89,edwards94}.
As noted later~\cite{lemaitre02c}, the essential feature that free-volume 
dynamics captures is the existence of an intensive quantity which measures 
internal disorder, and which evolves as the system orders or is driven away 
from equilibrium.
Alternative approaches characterize the internal state in terms of
an effective temperature~\cite{lemaitre03}.
The similarity between free-volume and a temperature results from 
the fact that, in hard-sphere materials at constant pressure, 
the dominant contribution to the energy of a subset of molecules is enthalpy.
In this case, local energy and density fluctuations are directly correlated.
Note finally, that the concept of Edwards' temperature is likely to be related 
to effective temperatures arising in weak versions of the fluctuation-dissipation 
theorem~\cite{cugliandolo97b,cugliandolo97c,ono02,makse02}.
In this paper, we  use free-volume terminology,
and refer the reader to~\cite{lemaitre03} where the relation between 
free-volume and effective temperature is  discussed in more detail.

\section{III. Constitutive equations}
\label{sec:model}

Our presentation of the model is broken down into four steps: (i) preliminaries
associated with the SFA, explicitly accounting for the finite stiffness, (ii) STZ theory, (iii) free volume equations, and 
(iv) rescaling to obtain dimensionless equations.
The equations in sections (ii-iv) have been presented elsewhere, 
but we include their derivation for the sake of completeness. 
We refer the reader to~\cite{falk98} and~\cite{lemaitre02c,lemaitre03}
for more detailed discussions of the underlying assumptions.

\subsection{(i) Surface Force Apparatus}

A primitive model of the SFA consists of a single slider block, pulled along a surface
by a spring of stiffness $k$. The opposite end of the spring  advances at a prescribed
velocity $V$.  Letting $x$ denote the displacement of the block relative to the stationary
surface, the spring exerts a force $F=k(Vt-x)$, where $t$ is time measured from some initial
time $t=0$ when $x=0$ .
The block  is subject to both the pulling force from the spring
and frictional resistance at the surface. Modeling the SFA as a single slider
assumes that the sliding surface is sufficiently small, rigid, and uniform and that
friction at the contact is sufficiently self-averaging that slip occurs  uniformly
across the interface. 

Assuming the thickness $h$
of the interface remains constant, the motion of the slider is related to the 
rate of shear deformation $\dot\gamma$ of the interfacial material by,
$\dot x = 2\,h\,\dot\gamma$.
If the area $S$ of the contact
is constant, the shear stress exerted by the slider
on the interfacial material can be written, $\sigma=F/S$.
Furthermore, for the experiments we  consider, the friction is sufficiently strong
that the motion is overdamped, and inertial forces  associated with the nonzero
mass of the slider can be neglected.
This leads to an equation of motion for the stress:
\begin{equation}
\dot\sigma = \mu\,(\dot\epsilon-\dot\gamma)
\label{eqn:stress}
\end{equation}
with, $\mu=h\,k/S$, and $\dot\epsilon=V/(2\,h)$.
The motion $\dot x(t)$ of the slider
follows from the solution to (1), which
depends  on  how
the strain-rate $\dot\gamma$ is related to
shear stress $\sigma$ and the internal state variables.

We next define the constitutive equations
which couple free-volume and STZ dynamics. 
Later we  restrict our discussion to subsets and (linearized)
simplifications of these equations 
to isolate effects associated with  distinct
state variables and/or nonlinearities, and to illustrate
phenomena  which require  a larger number of
state variables  to be resolved.

\subsection{(ii) Elements of STZ theory}
STZ theory is based on the  idea that the macroscopic deformation
of an amorphous material results from localized rearrangements 
involving cooperative molecular motion
at  mesoscopic scales~\cite{maeda78,maeda81,takeuchi87}.
The loci of such rearrangements are called shear transformation zones,
and the internal state of the system is characterized
by their number density.
In its simplest form, STZ theory involves only two types of zones
(labeled \lq\lq +" and  \lq\lq -", with number density 
$n_+$ and $n_-$, respectively)
oriented along the principal axes of the shear stress.
Zones of each types transform into one another during an elementary shear.
The average strain rate $\dot \gamma$  is given by
\begin{equation}
\label{eqn:gammadot}
\dot\gamma = {\cal A}_0\,(R_+\,n_+-R_-\,n_-)
\quad.
\end{equation}
Here $\dot \gamma$
averages over the populations of zones, $n_\pm$, 
that reorient with probabilities $R_\pm$, respectively.

The important insight Falk and Langer contributed to previous 
STZ theories 
was to treat the populations densities $n_\pm$ as state variables, 
and propose equations of motion for the populations. These take 
the following form~\cite{falk98,falk00}:
\begin{equation}
\label{eqn:npm}
\dot n_\pm = -R_\mp\,n_\mp + R_\pm\,n_\pm + \sigma\dot\gamma\,({\cal A}_c-{\cal A}_a\, n_\pm)
\quad.
\end{equation}
The first two terms on the right hand side
account for exchanges between the populations
of STZ's due to mesoscopic rearrangements, while the last term
introduces a coupling with the mean flow, through creation of STZ's at rate ${\cal A}_c$, and
annihilation at rate ${\cal A}_a$.  The equations 
describe how
shear deformations induce small displacements of the molecules,
hence creating and destroying shear transformation zones.
In this framework, the emergence of a yield stress in amorphous solids 
at low temperature is associated with the mobilization
of zones when  stress is applied. 


\subsection{(iii) Free-volume activation}

In the original formulation of STZ theory, the rates were estimated
to be nonlinear functions of stress. Derivation of the rates were based
on free-volume activation, as developed by Cohen, Turnbull and
coworkers to understand the phenomenology 
of the glass transition~\cite{turnbull61,turnbull70,grest81}.
Specifically, the rates $R_\pm$ were estimated to be of 
the form $\exp(-v_0(\pm\sigma)/v_f)$, 
where $v_0$ is a stress-dependent activation volume, and
$v_f$ is  a material dependent constant.
The detailed formulation of transformation rates $R_\pm$ is not 
essential to capture the STZ mechanism for jamming. Instead, 
a first order approximation for the stress dependence 
of the rates $R_\pm$ is sufficient~\cite{falk98,falk00}.

However, for hard-sphere systems, the free-volume $v_f$ is 
directly related to the density.
There is no reason to believe that it should take a fixed value
as a function of pressure and temperature.
On the contrary, it is a dynamical quantity which evolves as the material 
dilates or contracts. 
This observation
naturally
leads to dynamical equations for the free-volume, written in analogy 
with equations of motion for the populations $n_\pm$~\cite{lemaitre02b, lemaitre02c}.
While free-volume accounts for the existence of disorder in molecular packing, 
the difference in STZ densities describe its anisotropy.

Activation factors depends both on free-volume and stress fluctuations.
Assuming these effects are uncorrelated, we express the rates as:
\begin{equation}
R_\pm(\sigma,v_f) = R_0 \exp\left[-{v_0\over v_f}\right]\,\exp\left[\pm{\sigma\over\bar\mu}\right]
\quad.
\end{equation}
An elementary shear rearrangement takes a ``+'' oriented zone to a ``-'' oriented 
zone, or vice versa, and occurs if sufficient free-volume (large than $v_0$) 
is available, and if the virtual work of shear forces promotes the transition 
in the $\pm\to\mp$ direction.
The variable $\bar\mu$ is here a scale of forces (not to be confused with the elastic
modulus $\mu$) which may depend on temperature, and governs stress activation factors.
Note that the introduction of backward and forward jumps is an old idea, already present
in Eyring's theory of viscous liquids~\cite{eyring36} or in Spaepen's approach
to creep in metallic glasses~\cite{spaepen77}.

The equation of motion for $v_f$ is given by
\begin{equation}
\label{eqn:vf}
\dot v_f = -R_1\,\exp\left[-{v_1\over v_f}\right] + {\cal A}_v\,\sigma\,\dot\gamma\
\quad.
\end{equation}
Here free-volume dynamics involve two competing
mechanisms: 
(i) activated elementary compaction  which increase the density,
and
(ii) the transfer of macroscopic work into enthalpy, which
dilates the material.
The parameter ${\cal A}_v$ specifies how efficiently the work of external 
forces is used in dilatancy.
The parameter $R_1$ is an update frequency, which should be of the same order
as $R_0$; the activation volume $v_1$ may  differ from $v_0$, 
because the two elementary rearrangements (shear and compaction) 
involve different relative motion of the molecules, 
hence different reactional pathways. 
The ratio $\kappa=v_1/v_0$ is an essential parameter of the theory, and is expected 
to depend on the shape of the molecules.

The equations described here correspond to the low temperature limit of 
a more general set of equations for the dynamics of 
a disorder temperature~\cite{lemaitre03}.
We will restrict our current discussion to this 
free-volume formulation. As we will show, it 
captures a wide range of phenomena observed experimentally.

\subsection{(iv) Rescaling and change of variables}
\label{sec:lambda}

For the constitutive equations defined above 
(equations~(\ref{eqn:gammadot})~(\ref{eqn:npm})~(\ref{eqn:vf})),
it is convenient to introduce reduced variables~\cite{falk98,falk00},
\begin{equation}
\Delta = {n_--n_+\over n_\infty}\quad,\quad 
\Lambda = {n_-+n_+\over n_\infty}\quad,\quad{\rm and}\quad
\chi = {v_f\over v_0}
\end{equation}
along with the rescaled parameters, $n_\infty=2{\cal A}_c/{\cal A}_a$,
$\epsilon_0 = {{\cal A}_0\,{\cal A}_c/{\cal A}_a}$,
$\mu_0 = 1/({\cal A}_0\,{\cal A}_c)$, $E_0 = 2\epsilon_0\,R_0$, and $E_1=R_1/v_0$.
This change of variables leads to the following set of equations:
\begin{eqnarray}
\label{eqn:stzdil:1}
\dot\gamma&=&E_0\,\exp\left[{-{1\over\chi}}\right]\,
\left(\Lambda\,\sinh\left({\sigma\over\bar\mu}\right)-\Delta\,\cosh\left({\sigma\over\bar\mu}\right)\right)\\
\label{eqn:stzdil:2}
\dot\Delta&=&{\dot\gamma\over\epsilon_0}\,\left(1-{\sigma\over\mu_0}\,\Delta\right)\\
\label{eqn:stzdil:3}
\dot\Lambda&=&{\sigma\,\dot\gamma\over\mu_0\,\epsilon_0}\,\left(1-\Lambda\right)\\
\label{eqn:stzdil:4}
\dot\chi &=& -E_1 \exp\left[-{\kappa\over\chi}\right]+\alpha\sigma\dot\gamma
\end{eqnarray}
Equations~(\ref{eqn:stzdil:1}),~(\ref{eqn:stzdil:2}), and~(\ref{eqn:stzdil:3})
-- free-volume $\chi$ in (\ref{eqn:stzdil:4}) being held constant --
are almost identical to the original formulation of STZ theory 
by Falk and Langer~\cite{falk98,falk00}.
The variable $\Lambda$ accounts for the total density of STZ's.
The steady state value  $\Lambda=1$  is a stable fixed point.
In STZ theory, $\Lambda$ differs from $1$ only in the initial transient, 
and its initial value is expected to depend on the type of annealing performed
during the creation of a glass from a high temperature liquid.
Here we are not concerned with transient features associated
with these annealing-dependent initial values of $\Lambda$, 
and will assume $\Lambda=1$ throughout.  This
eliminates equation~(\ref{eqn:stzdil:3}).
The variable $\Delta$, is a normalized difference $\lq\lq +"$ and $\lq\lq -"$  STZ densities.
It accounts for the anisotropy of the molecular structure.
For a fixed applied stress $\sigma$ (and fixed free-volume, $\chi$),
equations~(\ref{eqn:stzdil:1}) and~(\ref{eqn:stzdil:2}) were shown previously to account 
for a transition between an elastic regime (jamming) 
and a visco plastic regime (flowing).
The transition occurs at a yield stress, $\sigma_y$,  satisfying
\begin{equation}
\tanh\left(\frac{\sigma_y}{\bar\mu}\right)=\frac{\mu_0}{\sigma_y}
\label{eqn:sigma_y}
\quad.
\end{equation}

The variable $\chi$ is a normalized free-volume, which accounts for 
the existence of disorder in the molecular structure.
Its dynamics, determined by equation~(\ref{eqn:stzdil:4}), couples
with equations~(\ref{eqn:stzdil:1}) and~(\ref{eqn:stzdil:2}) only as far as $\chi$
appears in the prefactor of equation~(\ref{eqn:stzdil:1}),
setting the time scale of elementary shear events.

At a fixed applied stress $\sigma$, equation~(\ref{eqn:stzdil:1}),~(\ref{eqn:stzdil:2})
and~(\ref{eqn:stzdil:4}) account for plastic deformation resulting from 
the coupled dynamics of $\Delta$ and $\chi$.
When the system is driven at a constant shear rate as in the SFA, these equations must be supplemented
with equation~(\ref{eqn:stress}), which accounts for coupling of material deformation
with a compliant driving apparatus.

\subsection{Discussion}

Most of the results presented here are consequences of free-volume 
dynamics~(\ref{eqn:stzdil:4}) coupled to equation~(\ref{eqn:stzdil:1}),
the STZ variables $\Delta$ (and $\Lambda$)
being held constant ($\Delta=0$ and $\Lambda=1$).
In this case, the equations~(\ref{eqn:stzdil:1}) and~(\ref{eqn:stzdil:4}) reduce to:
\begin{eqnarray}
\label{eqn:vf:1}
\dot\gamma &=& 
E_0\,\exp\left[-{1\over\chi}\right]\,\sinh\left[{\sigma\over\bar\mu}\right]\\
\label{eqn:vf:2}
\dot\chi &=& -E_1\,\exp\left[-{\kappa\over\chi}\right] + \alpha \,\sigma\,\dot\gamma
\end{eqnarray}
These equations can also be linearized for small stresses, which enables us to 
eliminate the parameter $\bar\mu$ (taken to unity) which enters the exponential
activation factors to fix a scale of stresses. The linearized (in $\sigma$) equations are
\begin{eqnarray}
\label{eqn:vflin:1}
\dot\gamma &=& E_0\,\exp\left[-{1\over\chi}\right]\,\sigma\\
\label{eqn:vflin:2}
\dot\chi &=& -E_1\,\exp\left[-{\kappa\over\chi}\right] + \alpha \,\sigma\,\dot\gamma
\end{eqnarray}
We refer to this as the {\it stress-linear} approximation. It
captures most of the phenomenology accompanying
the stick-slip instability, and allows for interesting analytical calculations.

Throughout this paper, we will use these three sets of equations-- the full, coupled system of equations
including STZ and free-volume effects, the nonlinear equations describing free volume dynamics only,
and the linearized version of the free-volume equations-- in order to 
clarify the consequences of our assumptions and the role of the different state variables.
The questions raised are: What behavior is already captured by free-volume dynamics, 
in the stress-linear version (equations~(\ref{eqn:vflin:1}-\ref{eqn:vflin:2}))? 
What is the importance of activation factors,
and the non-linear dependence of strain-rate versus stress 
(equations~(\ref{eqn:vf:1}-\ref{eqn:vf:2}))? 
What are the expected consequences of the interaction between 
several internal state variables,
as modeled by equations~(\ref{eqn:stzdil:1}),~(\ref{eqn:stzdil:2})
and,~(\ref{eqn:stzdil:4})?

\section{IV. Comparison with experiments}
\label{sec:comp}

In this section, we compare the qualitative behavior of our model with 
several important experimental observations.
Most of our investigation focuses on the equations governing free-volume dynamics,
either in their non-linear form~(\ref{eqn:vf:1}-\ref{eqn:vf:2})
or in their linear form~(\ref{eqn:vflin:1}-\ref{eqn:vflin:2}).
We study (i) stress versus strain rate relations in steady sliding,
(ii) transient dynamics upon start-up, and 
(iii) the transitions to stick-slip at low velocities.
In the last part of our work, we focus on the existence 
of (iv) chaotic regimes of stick-slip.
In order to observe chaos, we need the full (three-dimensional) 
set of non-linear equations~(\ref{eqn:stzdil:1}-\ref{eqn:stzdil:4}).

\subsection{(i) Steady sliding}

The first step in characterizing
the behavior of sheared materials is, of course,
the steady state relation between stress and strain rate.
In boundary lubrication and single asperity friction, 
it has been observed that, at low velocities, 
the friction force is weakly velocity weakening.
This weak dependence is reminiscent of the original observations 
by Dieterich~\cite{dieterich78,dieterich79}
in the case of dry surfaces, where a logarithmic dependence can be assumed
with a reasonable degree of confidence.
In boundary lubrication and single asperity friction there is insufficient data
to conclusively determine whether the 
steady state relation between force and velocity is 
logarithmic, power law, or a combination of the two~\cite{drummond00,bureau02}.
Below we calculate the stress vs. strain rate relationship in
order of increasing model complexity, beginning with the
linearized free-volume equations, followed by the nonlinear free volume equations,
and finally for the complete model, including STZ's.

\subsubsection{Stress-linear equations}
In the stress-linear version of our model
(equations~(\ref{eqn:vflin:1}) and~(\ref{eqn:vflin:2}))
the ratio between stress and strain rate determines a viscosity,
\begin{equation}
\eta = \exp(1/\chi)/E_0
\label{eqn:visc}
\end{equation}
which is a simple function of the 
free-volume $\chi$.

Initially, we take   the limit of infinite stiffness for the experimental apparatus, 
so that, from equation~(\ref{eqn:stress}), $\dot\gamma=\dot\epsilon$. 
>From equations~(\ref{eqn:vflin:1}) and~(\ref{eqn:vflin:2}), 
the dynamics of $\chi$ reduces to:
\begin{equation}
\dot\chi = -E_1\,\exp\left[-{\kappa\over\chi}\right] 
+ {\alpha\over E_0}\,\exp\left[{1\over\chi}\right]\,\dot\epsilon^2
\quad.
\end{equation}
The quantity $\chi$ admits a steady state value only if the shear rate
is not too large. For high shear rates,
\begin{equation}
\label{eqn:epsilonstar}
\dot\epsilon > \dot\epsilon^* = \sqrt{E_0\,E_1\over\alpha}
\quad,
\end{equation}
this equation becomes unstable, $\dot\chi$ is positive at all times, 
and $\chi$ diverges. This instability is not related to stick-slip,
since it occurs for any value of the stiffness. It indicates that at large 
shear rates, the material cannot dissipate the work of external forces, 
and is driven towards a highly disordered state.
Situations when $\dot\epsilon>\dot\epsilon^*$ are by definition transient: 
the material does not reach a steady state. Instead $\chi$ diverges.

Interpreting $\chi$ as a free-volume, this divergence may correspond
to an opening of the interface, and possible loss of contact between the surfaces.
In typical experiments, if such high shear rates are applied,
the divergence of $\chi$ would be  {\it a priori}
limited by 
the scanning length of the experimental device.
It is intriguing, however, that opening modes at large shear rates are observed
experimentally~\cite{delphine}. They have not been characterized in
the literature because they are usually viewed as experimental technicalities
which limit the range of accessible driving velocities.
Instead, experiments have been restricted to situations in which such effects do not occur,
specifically
low driving shear 
rates, $\dot\epsilon<\dot\epsilon^*$, or high shear rates on limited 
scanning lengths. 
In the latter case, the apparent viscosity, 
$\eta = \exp(1/\chi)/E_0\sim 1/E_0$ is approximately independent of $\chi$, and
it may appear, from a measurement of stress alone, 
that the system is stationary  although the 
internal dynamics may not have reached a steady state.

The existence of this divergence at a critical driving shear rate 
$\dot\epsilon^*$ 
arises physically from  the fact that the transition probabilities are bounded.
The factor $\exp(1/\chi)$ approaches unity  at high $\chi$,  and becomes decreasingly sensitive
to changes in the intensive variable $\chi$ when $\chi$ 
is much larger than 1.
At this point, free-volume dynamics is only weakly  coupled to the other equations.
Of course, if a high shear rate $\dot\epsilon>\dot\epsilon^*$ is applied steadily,
this divergence ultimately leads (in the $t\to\infty$ limit)
the system towards a highly disordered, ``fluidized'', state.
Our equations are not designed to describe this limit, and instead should start to 
break down when $\chi$ reaches some large values, say $\chi_f$.
We expect $\chi_f>>1$ (here the value 1 corresponds 
to the activation free-volume for an elementary shear transformation,
which is much less than one molecular volume per molecule,
whereas $\chi_f$ is of the order of one additional molecular volume of free space per molecule).
We do not attempt to account for the late stages of this divergence,
and only changes of material behavior around $\chi\sim1$ are of interest to us.
For this reason, and to simplify the discussion, we can safely take $\chi_f$ to infinity.

We could  avoid explicit reference to such a divergence as several
mechanisms could be invoked to explain the saturation of free-volume at high
velocity resulting in the emergence of another fixed point. Possible mechanisms include: 
weak density dependencies of various parameters, like $E_0$, $E_1$, or $\alpha$,
or higher order terms in the free-volume dynamics 
(equation~(\ref{eqn:stzdil:4}),~(\ref{eqn:vf:2}), or~(\ref{eqn:vflin:2})).
Here we study free-volume dynamics in the simplest form for several reasons.
Firstly, several mechanisms could be invoked to account for the
saturation of free-volume at high velocity, and it is unclear at this stage which would be
dominant. Thus we prefer not to differentiate between them.
Secondly, regardless of the mechanism, the approach to any high free-volume fixed point would, 
in its early stages, closely resemble  the divergence we discuss here.
Given the limited span of experiments, we expect our model 
is relevant to observations which are currently available.

For shear rates smaller than $\dot\epsilon^*$, 
free-volume admits a steady state value, and this leads to a relation 
between stress and strain rate, of the form:
\begin{equation}
\sigma= {\dot\epsilon^{n}\over E_0}\,\left({\alpha\over E_0 E_1}\right)^{{n-1\over2}}
\quad.
\label{eqn:linss}
\end{equation}
The constitutive equations thus account for power-law viscosity,
with an exponent
\begin{equation}
n={\kappa-1\over\kappa+1}
\label{eqn:n}
\end{equation}
which is directly related to the ratio $\kappa=v_1/v_0$ of activation volumes.
For $\kappa>1$ the stress is an increasing function of the shear rate.
The material is shear thickening. 
For $\kappa<1$, the stress is a decreasing function of the shear rate,
the material is shear thinning.
We will see that, in the later case, the system exhibits a transition to 
stick-slip at low velocities. 

\subsubsection{Non-linear free-volume equations}

Next we calculate the stress vs. strain rate for the nonlinear free volume 
equations~(\ref{eqn:vf:1}) and~(\ref{eqn:vf:2}). In this case, we obtain a
generalization of Eq.~(\ref{eqn:linss}) which is valid beyond the range 
of small stresses, where the linear stress approximation is expected to hold.
The resulting relationship 
is given by:
\begin{equation}
\dot\epsilon = E_0\,\left({\alpha E_0\over E_1}\right)^{{1\over\kappa -1}} 
\,\sigma^{{1\over\kappa -1}}\,\left(\sinh\left({\sigma\over\bar\mu}\right)\right)^{{\kappa\over\kappa -1}} 
\quad.
\label{eqn:nlss}
\end{equation}
For small $\sigma$ linearization of the hyperbolic sine, leads to the power law
rheology  described above (Eqs.~(\ref{eqn:linss}--\ref{eqn:n})).
For large stresses, the right hand side is dominated by the exponential growth
in the hyperbolic sine,
which results in a logarithmic dependence of the stress $\sigma$ on strain rate $\dot\epsilon$.
This expression is compatible with experimental observations, although
as already mentioned, the range of available  observations is sufficiently limited to render
detailed functional fits
inconclusive~\cite{drummond00,bureau02,gourdon03}.

As in the linearized case, Eq.~(\ref{eqn:nlss}) is accompanied by a condition imposed on the strain
rate  $\dot\epsilon$ which must be satisfied for there to be a self-consistent, steady state value of $\chi$.
Violation of the condition ($\dot\epsilon>\dot\epsilon^*$) is 
associated with diverging free volume and fluidization.
For the nonlinear free volume equations
the constraint is more complicated than in the linearized case, and is described by
$\dot\epsilon/(E_0 \sinh(\sigma/\bar\mu))=\exp(-1/\chi)<1$. 
A stationary value of $\chi$ exists if and only if the inequality is satisfied. Equivalently,
a solution exists iff:
\begin{equation}
\left({\alpha E_0\over E_1}\,\sigma\,\sinh\left({\sigma\over\bar\mu}\right)\right)^{{1\over\kappa -1}} <1
\quad.
\label{eqn:divergence}
\end{equation}
The case of equality in Eq.~(\ref{eqn:divergence}) can be used to define a critical value $\sigma^*$:
\begin{equation}
{\alpha E_0\over E_1}\,\sigma^*\,\sinh\left({\sigma^*\over\bar\mu}\right)=1
\quad.
\end{equation}
Then, validity of Eq.~(\ref{eqn:divergence}) depends on
the value of $\sigma$ relative to $\sigma^*$.
If $\kappa<1$, $\sigma$ is a decreasing function of $\dot\epsilon$, and the 
condition~(\ref{eqn:divergence}) is met for $\sigma>\sigma^*$, 
or equivalently, $\dot\epsilon<\dot\epsilon^*=\dot\epsilon(\sigma^*)$.
If $\kappa>1$, $\sigma$ is an increasing function of $\dot\epsilon$, and the condition~(\ref{eqn:divergence}) is met for $\sigma<\sigma^*$, or $\dot\epsilon<\dot\epsilon^*$.
In both cases, as with the linear equations, 
there is a limiting driving strain rate, $\dot\epsilon^*$ above which
the material fluidizes, and (in our model) the high-velocity fixed point is 
sent to infinity.
For $\dot\epsilon>\dot\epsilon^*$, as the quantity $\chi$ diverges, the 
relation between stress and strain rate reduces to
\begin{equation}
\dot\epsilon=E_0\,\sinh\left({\sigma\over\bar\mu}\right)
\quad.
\end{equation}
Stress appears to be increasing for large values of the strain rate.

\subsubsection{Full Free Volume and STZ equations}
Finally, we explore the steady state stress vs. strain rate relationship using the 
complete set of 
equations~(\ref{eqn:stzdil:1}),~(\ref{eqn:stzdil:2}),~(\ref{eqn:stzdil:3}), and~(\ref{eqn:stzdil:4}). In
this case, the steady state relation between stress and strain rate reads:
\begin{equation}
\label{eqn:stzdil:stat}
\dot\epsilon = E_0\,\left({\alpha E_0\,\sigma\over E_1}\right)^{{1\over\kappa -1}} 
\,\left(\sinh\left({\sigma\over\bar\mu}\right)-\frac{\mu_0}{\sigma}\,\cosh\left({\sigma\over\bar\mu}\right)\right)^{{\kappa\over\kappa -1}} 
\quad.
\end{equation}
The complete equations capture the  physical phenomena of jamming (i.e. the absence of
flow), which occurs 
for stresses smaller than a yield stress, $\sigma_y$, which satisfies equation (\ref{eqn:sigma_y}).
Otherwise, the behavior is very similar to our previous results for 
the nonlinear free volume equations.
For non-vanishing shear rates, we can again define $\sigma^*>\sigma_y$
as the solution of,
\begin{equation}
{\alpha E_0\over E_1}\,
\left(\sigma^*\,\sinh\left({\sigma^*\over\bar\mu}\right)
-\mu_0\cosh\left({\sigma^*\over\bar\mu}\right)\right)=1
\end{equation}
and $\dot\epsilon^*=\dot\epsilon(\sigma^*)$.
Again, for $\kappa<1$, the stress vs. strain rate relation is decreasing,
and for $\kappa>1$, it is increasing. 
Again, the range of strain rates where steady sliding is reached is bounded 
by $\dot\epsilon<\dot\epsilon^*$.
We see from equation~(\ref{eqn:stzdil:stat}) that the STZ dynamics do not  significantly change the sliding properties
as soon as the relevant values of the stress are large compared to $\sigma_y$.
This situation is particularly relevant for stick-slip instabilities, which
occur for $\kappa<1$, in the low velocity, large stress regime.

\subsection{(ii) Transient Dynamics Characterizing the Approach to Steady Sliding}

Next we describe transient effects associated with discrete jumps in the applied strain rate.
Both the initialization of the system, as well as several common experimental procedures probing
transients under controlled conditions, can be described in this general framework. 
We first consider the initial, waiting time dependent transients associated with starting
the system from rest. This is followed by an examination of the experimental protocol,
referred to as stop-start, or slide-hold-slide tests, in which the system is prepared in the steady
state before the waiting time begins.
For this analysis, we will focus on the linear and nonlinear free volume equations.  
Inclusion of the STZ terms does not quantitatively alter the results.

\subsubsection{Transient dynamics upon start-up}

In the absence of forcing, $\sigma=0$, the linear and nonlinear free volume
equations are equivalent.
>From equation~(\ref{eqn:stzdil:4}) (or~(\ref{eqn:vf:2}), or~(\ref{eqn:vflin:2})),
the free volume  $\chi$ relaxes to $0$.
The late stages of this relaxation (as the waiting time, $t_w$, goes to $\infty$) 
are logarithmic in time:
\begin{equation}
\chi(t_w) \simeq {\kappa\over \log(E_1 t_w)}
\quad.
\end{equation}
Simultaneously, as the system becomes increasingly compact,
the effective ``viscosity'' (Eq.~(\ref{eqn:visc})) increases:
\begin{equation}
\eta(t_w) \equiv {1\over E_0}\,\exp\left[{1\over\chi}\right]\,
\simeq {1\over E_0}\,(E_1 t_w)^{1/\kappa}
 \quad.
\end{equation}
The immediate consequence of this time dependent relaxation is that the material displays
an age dependent initial value of the stress when shear is subsequently applied.

\begin{figure}
\begin{center}
\unitlength = 0.005\textwidth
\begin{picture}(100,110)(5,0)
\put(10,5){\resizebox{90\unitlength}{!}{\includegraphics{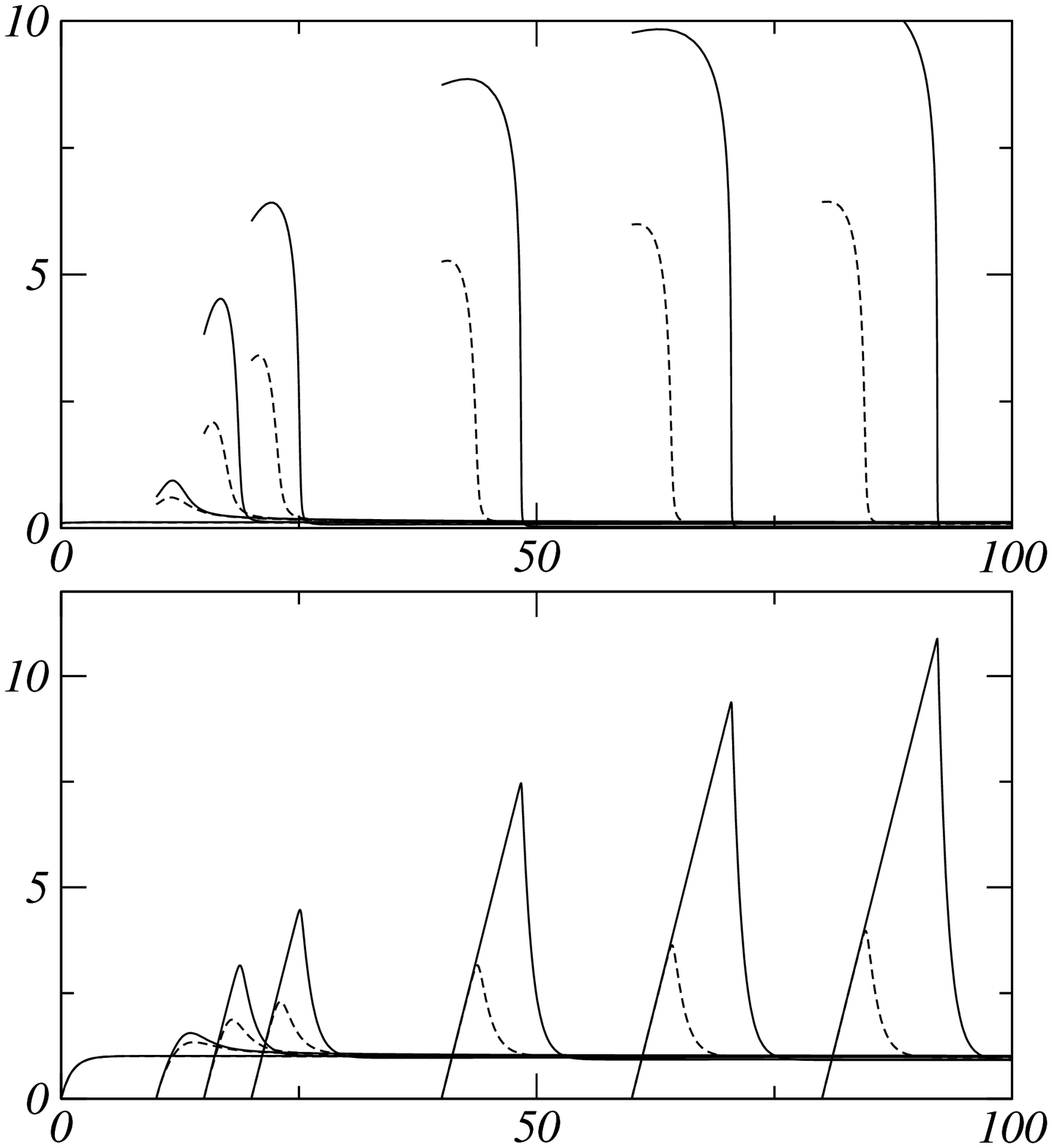}}}
\put(5,95){\makebox(0,0){\large $1/\chi$}}
\put(5,50){\makebox(0,0){\large $\sigma$}}
\put(90,2){\makebox(0,0){\large $t$}}
\end{picture}
\end{center}
\caption{\label{fig:bumps} Transient Dynamics upon start up from the integration 
of equation~(\ref{eqn:stress}) coupled to~(\ref{eqn:vf:1}) and~(\ref{eqn:vf:2}),
with $E_0=E_1=\alpha=1$. From equation~(\ref{eqn:epsilonstar}), $\dot\epsilon^*=1$,
and the applied strain rate is $\dot\epsilon=0.9$.
Results for $\kappa=0.8$ (solid lines) and $\kappa=1.2$ (dashed lines) are shown.
No stress is applied during the initial density relaxation of the material  from $t=0$
to $t=t_w$, marked by the increase of stress.
At $t_w$ the strain rate is suddenly applied and the ensuing
dynamics of $1/\chi$ (top) and  $\sigma$ (bottom) are displayed.
Here smaller values of  $\kappa$ (i.e. $\kappa=0.8$) leads to a larger value of 
the dynamical yield strength,  
which results from the fact that a smaller value of $\chi$ (more compact state) is reached
is reached during the waiting period.
}
\end{figure}

This is illustrated
in Figure~\ref{fig:bumps}  where we plot
$\chi$ and $\sigma$ as a function of time, obtained
from the integration 
of equation~(\ref{eqn:stress}) coupled to~(\ref{eqn:vf:1}) and~(\ref{eqn:vf:2}).
Distinct peaks correspond to samples of different ages at the onset of the applied strain rate.
In every case, the system is initialized at $t=0$ with a large value ($\chi(t=0)=10$) of free-volume.
>From time $t=0$ to $t=t_w$, free-volume relaxes in the absence of applied shear.
At  time $t=t_w$, a strain rate $\dot\epsilon$ is suddenly applied, and held constant from
that time on.
For each curve, the value of $t_w$  corresponds to
the time when stress begins to ramp up.

The initial increase of the stress is due to the fact that $\chi$ has relaxed to a
relatively small value during the waiting time,  resulting in a high effecting viscosity, which
resists rapid deformation of the material.
When the stress gets large, free-volume increases. The system dilates, and
the shear deformation rate increases suddenly, resulting in the observed 
stress drop.

\subsubsection{Stiff apparatus}

To provide analytical estimates 
of the age dependence of the peak stress, we take the limit of an infinitely  stiff apparatus,
$\mu\to\infty$.
When a fixed shear rate is imposed after some waiting time $t_w$,
the interfacial material resists with a force
which is an increasing function of age $t_w$. 
For the linearized
equation~(\ref{eqn:vflin:1}), 
the stress upon start-up reads,
\begin{equation}
\sigma_{\rm start}(t_w)={1\over E_0}\,(E_1 t_w)^{1/\kappa}\,\dot\epsilon
\end{equation}
The response is viscous, with an instantaneous initial viscosity which 
increases as a power of waiting time. 
 
For the nonlinear free-volume equations~(\ref{eqn:vf:1}) the power law
dependence is replaced by a logarithm. 
For large $t_w$, the force upon start-up is:
\begin{equation}
\sigma_{\rm start}(t_w)
\simeq {\bar\mu\over\kappa}\,\log\left(E_1 t_w\right) 
+ \bar\mu\,\log\left({\dot\epsilon\over E_0}\right) 
\quad.
\end{equation}
This logarithmic dependence on waiting time and shear rate at long times 
is consistent with experimental observations,
although it must be noted that experiments performed with 
the SFA~\cite{drummond00,gourdon03} 
are rather inconclusive and that most significant evidence comes from the single asperity
experiments by Bureau {\it etal\/}~\cite{bureau02}.
Here, the logarithmic dependence is a direct consequence of the assumption 
that transformation rates $R_\pm$ depend exponentially on stress (i.e. proportional
to  $\exp(\pm\sigma/\bar\mu)$).

\subsubsection{Stop-start tests}
The characteristics of glassy materials depends
sensitively on sample age and preparation methods.
It is therefore important to focus on experimental protocols in which
the initial state can be relatively well-defined.
One common experimental convention which aims to control the initial state involves starting
from steady sliding motion. The initial state of the material is 
determined by the driving velocity.
In velocity step experiments 
the drive  velocity undergoes discrete changes from one value to another, 
and the transient response is monitored. This was one of the original protocols used
to investigate the correspondence between rate and state laws and 
dry friction experiments~\cite{dieterich78,dieterich79}, and was recently
investigated for lubricated contacts~\cite{drummond00}.
Another protocol, referred to as stop-start or slide-hold-slide experiments,
involves preparing the system in a constant velocity steady state, then
suddenly stopping the drive and letting the system relax for a time $t_w$,
and finally restarting at the initial velocity. This latter protocol has been 
studied extensively in boundary lubrication~\cite{gee90,yoshizawa93b,drummond01},
and for the plastic response
of single asperities at the contact between rough surfaces~\cite{bureau02}.

\begin{figure}
\begin{center}
\unitlength = 0.005\textwidth
\begin{picture}(100,110)(5,0)
\put(10,5){\resizebox{90\unitlength}{!}{\includegraphics{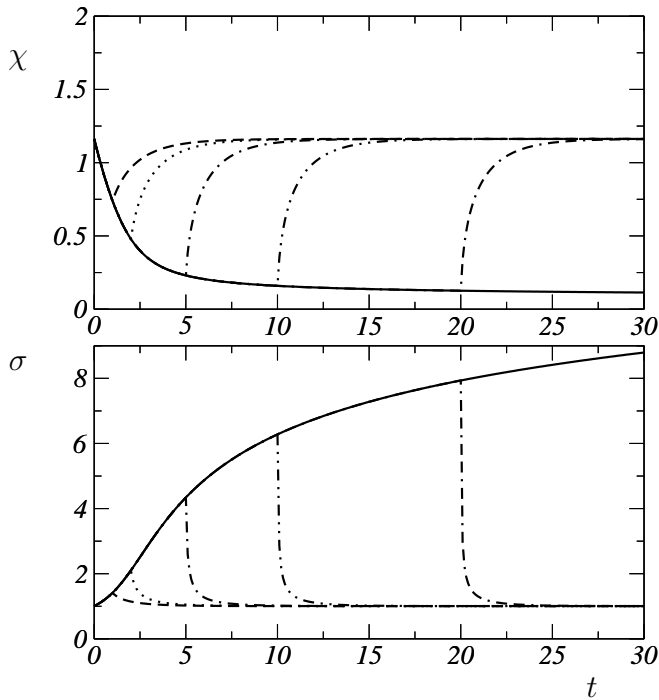}}}
\put(5,95){\makebox(0,0){\large $\chi$}}
\put(5,50){\makebox(0,0){\large $\sigma$}}
\put(90,2){\makebox(0,0){\large $t$}}
\end{picture}
\end{center}
\caption{\label{fig:stiff}
Numerical integration of equations~(\ref{eqn:vf:1}) 
and~(\ref{eqn:vf:2}) during stop-start tests for shear rate 
$\dot\gamma=\dot\epsilon=0.5$.
Parameters are $E_0=E_1=\alpha=\mu=1$, and $\kappa=0.8$.
For each curve, the applied shear haults for a time $t_w$, after which it
is  suddenly reapplied.  The
dynamics of $\chi$ (top) and  $\sigma$ (bottom) are displayed.
The solid lines indicate  the relaxation of $\chi$ in the absence
of shear, and the value of $\sigma$ upon start-up.
Different line styles are used for different $t_w$.
}
\end{figure}

This stop-start protocol 
grants direct access to the aging process from a well-controlled initial state.
Stop-start tests are depicted Figure~\ref{fig:stiff},
based on numerical integration
of equations~(\ref{eqn:vf:1}) and~(\ref{eqn:vf:2})
in the limit of an infinitely stiff apparatus. 
Here $\chi$ and $\sigma$ are plotted as functions of time. 
The initial value of $\chi$ corresponds to the steady state at a given
shear rate $\dot\epsilon=0.5$.
When the external drive haults (the \lq\lq stop" phase) $\chi$ relaxes to a smaller
value, as the lubricant becomes more compact.
The same shear rate in then suddenly reapplied (the \lq\lq start" phase) after different waiting times. 
In this $\mu\to\infty$ limit, $\dot\gamma=\dot\epsilon$, 
and the stress immediately takes  its peak value, where
the position of the peak marks the waiting time for each curve.
After start-up, free-volume increases, due to the transfer of 
the work of shear forces into enthalpy. This results in a decrease of the viscosity $\eta$,
accompanied by a decay of the stress with time.

\begin{figure}
\begin{center}
\unitlength = 0.005\textwidth
\begin{picture}(100,110)(5,0)
\put(10,5){\resizebox{90\unitlength}{!}{\includegraphics{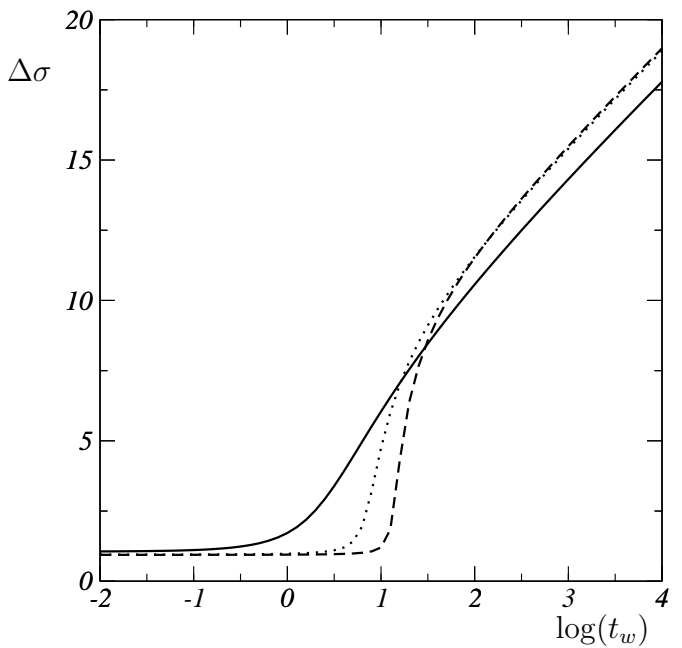}}}
\put(5,85){\makebox(0,0){\large $\Delta\sigma$}}
\put(90,2){\makebox(0,0){\large $\log(t_w)$}}
\end{picture}
\end{center}
\caption{\label{fig:latency}
Peak stress, $\Delta\sigma$ as a function of waiting time $t_w$,
from integration of equations~(\ref{eqn:vf:1}) and~(\ref{eqn:vf:2}).
Parameters are $E_0=E_1=\alpha=\mu=1$, and $\kappa=0.8$,
whence $\dot\epsilon^*=1$. 
Different shear rates have been used:
$\dot\gamma=\dot\epsilon=0.3$ (solid line), $0.9$ (dotted line) 
and $0.99$ (dashed line).
Approaching $\dot\epsilon^*=1$, a latency time interval becomes increasingly well defined,
indicating the increasing importance of free-volume dynamics beyond the
small-$\chi$ domain, where log-time relaxation of free-volume holds.
}
\end{figure}

The solid line in Figure~\ref{fig:stiff} marks the envelope of all response peaks, 
and thus defines the peak  stress as a function of the resting time $t_w$. 
Experimentally, for short waiting times, Yoshizawa {\it et al.}~\cite{yoshizawa93b}.
observed that for certain lubricants, the corresponding curve exhibits a well defined latency time.
That is, there is a threshold in the waiting time, below which no increase in the stress is observed.
For longer waiting times, Drummond {\it et al.} and 
Gourdon {\it et al.}~\cite{drummond00,gourdon03} found that the difference $\Delta\sigma$
between the peak value of the stress, and the steady state value (the so-called \lq\lq stiction spike")
increases as $\Delta\sigma\sim \log(t_w)$ for large $t_w$.
Both the short and
long time behavior is reproduced by the nonlinear free volume equations, as shown in
Figure~\ref{fig:latency} for different values
of the driving velocity. 
Note the latency time (roughly associated with the rapid rise of  $\Delta\sigma$) 
becomes increasingly sharply defined  at high drive velocities.
The follows from the fact that
at high velocities, approaching  the limiting shear rate $\dot\epsilon^*$,
the stationary value of free-volume is large, hence
free-volume activation factors (which scale as $\exp(1/\chi)$) are essentially constant.
Therefore, the free volumthe free volume $\chi$ nearly decouples from 
the relation between stress and strain rate.
If the external drive is stopped during a short time interval, $\chi$ 
relaxes, but the changes in $\chi$ have little effect on the dynamic viscosity $\eta$.
Therefore, when the shear is applied again, the relation between stress and strain
rate is still very close to steady state, and only a very faint peak is observed.

Note finally that, due to the small spatial extent over which shear can take place
in the SFA, it is likely that
in some experiments the internal state of the material 
never reaches the  steady state value.
It might appear that a steady state has been reached when
$\chi$ is large enough to almost decouple from the stress {\it vs.} strain-rate 
relation.
In this case, the latency time depends on the time it takes for $\chi$ to relax
to values of order unity, which will depend on the details of sample preparation.
This subtle history dependence may explain large fluctuations observed in
some experiments~\cite{yoshizawa93b,drummond00,drummond01}.

\subsection{(iii) Instability and transition to stick-slip}

Here, we show that our constitutive equations not only account for the 
existence of stick-slip behavior at low drive velocity, but that 
they reproduce the shape of the stick-slip cycle with remarkable accuracy.
They also account for the existence of continuous and discontinuous  
transitions to stick-slip depending on the stiffness of the apparatus, and
in some cases, chaos in the neighborhood of the transition.
For this analysis, the nonlinear free volume equations will be sufficient to 
characterize the transition from stick-slip to steady sliding, and 
the nature of the bifurcation. The full set of coupled constitutive equations,
including STZ, preserve the phase boundaries described by the free volume equations.
However, STZ effects (particularly, which introduce another dimension to the dynamical
system) are required to capture chaotic phenomena.

\subsubsection{Stick-slip motion: shape of the pulse}

\begin{figure}
\begin{center}
\unitlength = 0.005\textwidth
\begin{picture}(100,110)(5,0)
\put(10,5){\resizebox{90\unitlength}{!}{\includegraphics{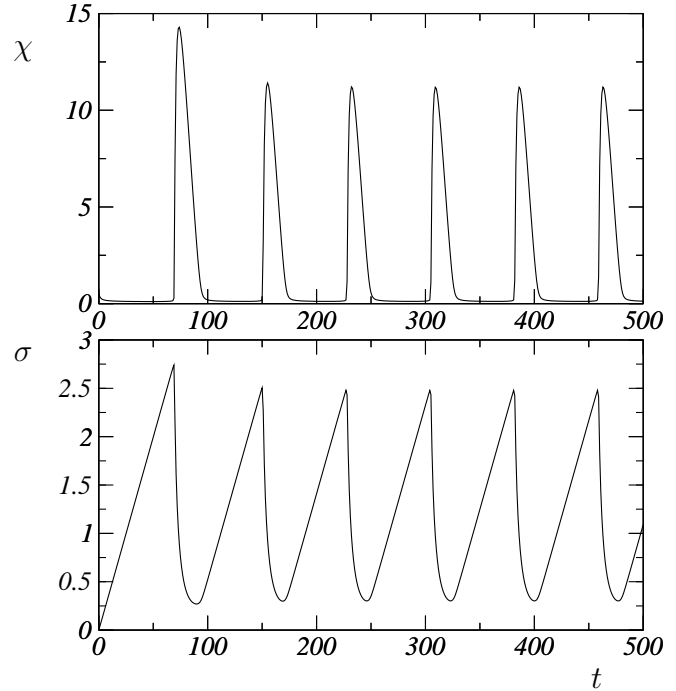}}}
\put(5,95){\makebox(0,0){\large $\chi$}}
\put(5,50){\makebox(0,0){\large $\sigma$}}
\put(90,2){\makebox(0,0){\large $t$}}
\end{picture}
\end{center}
\caption{ \label{fig:stickslip}
Stick-slip dynamics obtained from 
numerical integration of equations~(\ref{eqn:stress}),~(\ref{eqn:vf:1})
and~(\ref{eqn:vf:2}) for a fixed strain rate $\dot\epsilon=0.2$, and 
stiffness $\mu=0.2$.
Parameters are $E_0=E_1=\bar\mu=\alpha=1$, and $\kappa=0.8$.
The initial value of the free-volume is $\chi=1$.
The regime of steady plastic deformation is unstable and leads to stick-slip motion.
Fast relaxations of the stress result from sudden dilatancy of the material.
}
\end{figure}

At low drive velocities,  and for a sufficiently compliant apparatus, 
rather than sliding at constant velocity, the interface exhibits stick-slip motion.
We begin our analysis of this motion by illustrating a typical stick-slip cycle obtained from 
numerical integration of equation~(\ref{eqn:stress}) coupled 
with the constitutive equations~(\ref{eqn:vf:1}) and (\ref{eqn:vf:2}).
In Figure~\ref{fig:stickslip} we plot the  time series of $\chi$ and $\sigma$ during 
stick-slip.
As in the experiments, stick-slip cycles appear qualitatively similar to
stiction peaks. However, stick-slip arises when the driving motion is constant,
unlike stiction peaks in stop-start test,
which represent transient responses to
time varying slide-hold-slide drivers and are typically monitored
in regime where the steady state motion corresponds to constant velocity sliding,
Stick-slip arises due to an internal instability of the material at a given shear
rate. Rather than sliding at  that steady rate, the material alternates between \lq\lq sticking" 
and \lq\lq slipping." During the stick phase
free-volume decreases to a value which is small enough to hinder the relative displacement of 
lubricated surfaces, during which time the material creeps at a rate which is too slow to keep up with
the external drive. Consequently, the  stress builds up.
When it becomes large enough to trigger dilatancy, the slip phase begins, and $\chi$ 
increases suddenly, and the stress is released during rapid sliding motion.

\begin{figure}
\begin{center}
\unitlength = 0.005\textwidth
\begin{picture}(100,110)(5,0)
\put(10,5){\resizebox{90\unitlength}{!}{\includegraphics{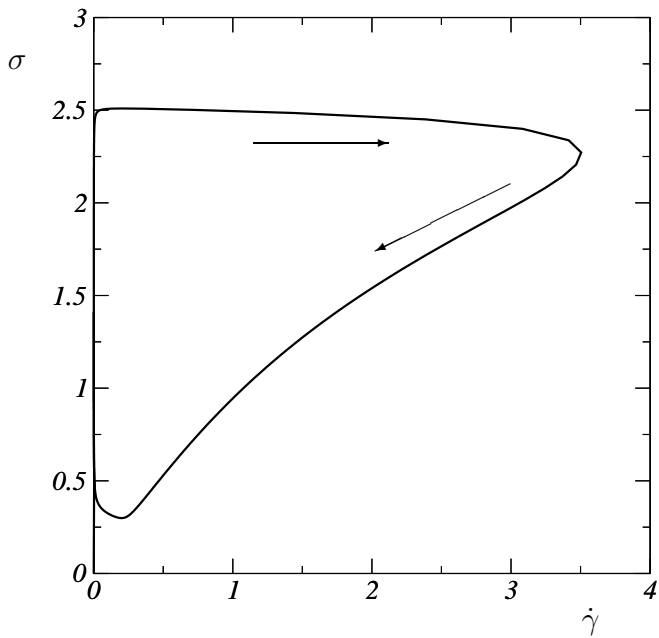}}}
\put(5,85){\makebox(0,0){\large $\sigma$}}
\put(50,75){\makebox(0,0){\vector(1,0){20}}}
\put(70,62){\makebox(0,0){\vector(-2,-1){20}}}
\put(90,2){\makebox(0,0){\large $\dot\gamma$}}
\end{picture}
\end{center}
\caption{ \label{fig:cycle}
Stick-slip cycle from the data of Figure~\ref{fig:stickslip}.
Arrows indicate the direction in which the stick-slip cycle is followed.
The ramps on Figure~\ref{fig:stickslip} correspond here to the increase of
$\sigma$ at vanishing $\dot\gamma$. The plastic strain rate suddenly increases
at almost constant $\sigma$ before the friction force starts to decrease.
}
\end{figure}

In Figure~\ref{fig:cycle} the same data is represented in a plot of 
$\sigma$ versus $\dot\gamma$, which compares  quite favorably
with typical cycles observed by Drummond and Israelachvili~\cite{drummond00}.

\subsubsection{Characterization of the Hopf Bifurcation}

\begin{figure}
\begin{center}
\unitlength = 0.005\textwidth
\begin{picture}(100,110)(5,0)
\put(10,5){\resizebox{90\unitlength}{!}{\includegraphics{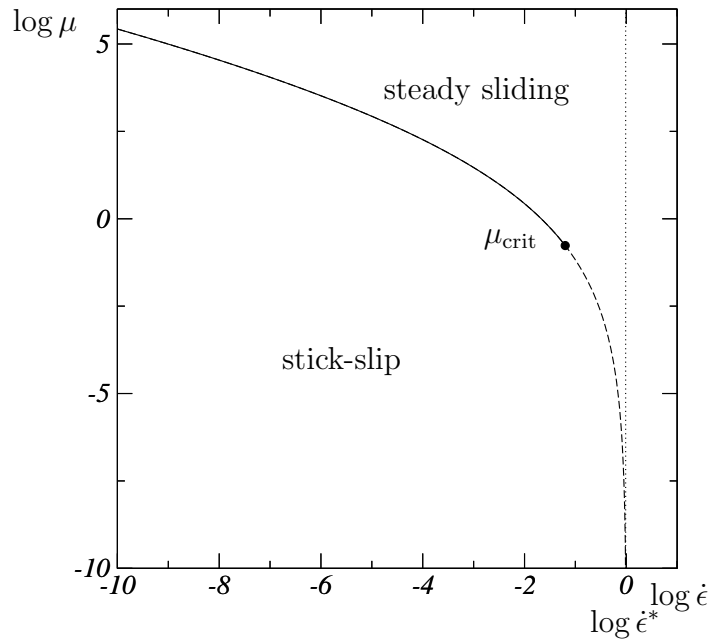}}}
\put(6,90){\makebox(0,0){\large$\log\mu$}}
\put(75,58){\makebox(0,0){\large$\mu_{\rm crit}$}}
\put(92,1){\makebox(0,0){\large$\log\dot\epsilon^*$}}
\put(100,5){\makebox(0,0){\large$\log\dot\epsilon$}}
\put(50,40){\makebox(0,0){\large stick-slip}}
\put(70,80){\makebox(0,0){\large steady sliding}}
\end{picture}
\end{center}
\caption{\label{fig:hopf} 
Phase diagram for equations~(\ref{eqn:stress}),~(\ref{eqn:vflin:1}) 
and~(\ref{eqn:vflin:2}), and $\kappa<1$, in $\log\dot\epsilon$-$\log\mu$ plane.
The dotted line indicates the limit velocity $\dot\epsilon^*$ beyond
which $\chi$ diverges.
The solid and dashed lines denote the curve $\mu_{\rm hopf}$ below which
steady sliding motion is unstable; $\mu_{\rm hopf}$ vanishes at
$\dot\epsilon=\dot\epsilon^*$.
The solid part of this line corresponds to points where the transition 
to stick-slip is continuous (super-critical) while the dashed line corresponds 
to points where the transition is discontinuous (sub-critical).
When the transition is discontinuous, there is a zone of bistability,
which lies above the dashed line.
}
\end{figure}

With decreasing velocity and for a compliant apparatus,
our constitutive equations exhibit a Hopf 
bifurcation separating steady sliding from  stick-slip dynamics. 
The locus of bifurcation points
defines a phase  boundary in the $\mu$ (stiffness) vs. $\dot\epsilon$ (strain rate) plane.
A systematic analysis of the emergence of stick-slip motion is easily performed
for the stress-linear version of the constitutive equations. 
The details are given in Appendix I.
The critical stiffness defining the Hopf bifurcation point 
for Eqs.~(\ref{eqn:stress}),~(\ref{eqn:vf:1}) 
and~(\ref{eqn:vf:2})  is given by
\begin{equation}
\mu_{\rm hopf} = {E_1\over E_0}\,{1-\kappa\over(\kappa+1)^2}\;
\left({\alpha\,\dot\epsilon^2\over E_0\,E_1}\right)^{{\kappa-1\over\kappa+1}}\;
\ln\left[{\alpha\,\dot\epsilon^2\over E_0\,E_1}\right]^2
\quad.
\end{equation}
Solving  for
$\mu$ as a function of $\dot\epsilon$  with all other parameters held fixed, defines the
phase boundary 
in the $(\mu,\dot\epsilon)$ plane, below which steady sliding becomes 
unstable (see Figure~(\ref{fig:hopf})). 
Our curve is  exhibits the same convexity as available 
experimental data~\cite{drummond01,baumberger94}. Additional experiments are
needed to provide a quantitative experimental characterization of the curve.

Experimental data also suggest that different types of transitions
between stick-slip and steady sliding are possible. The transition may be 
continuous (approaching the phase boundary the amplitude of stick slip motion 
decreases continuously to zero as drive
velocity is increased) or discontinuous (there is an abrupt
change from finite amplitude stick-slip spikes to steady sliding as drive velocity is increased).
Continuous transitions correspond to supercritical Hopf bifurcations, and discontinuous transitions
correspond to subcritical bifurcations. The later case is typically accompanied by hysteresis (i.e. 
coexistence of stick-slip and steady sliding, such that a steady decrease in the drive velocity 
results in a transition to stick slip at a lower strain rate than that associated with 
the transition to steady sliding
associated with increasing velocity from
the stick-slip phase).
This feature is observed in both lubricated friction~\cite{drummond01}
as well as dry interfaces. in the latter case  it 
is better characterized~\cite{baumberger94}, though the physical
origin of the similarity (if indeed it persists upon more detailed
experimental investigations) remains unclear. 
For both cases, to date analytical models have failed to capture the existence
of both super- and sub-critical Hopf  bifurcations.

The transition point separating super- and subcritical Hopf bifurcations on the phase boundary
in the $(\mu,\dot\epsilon)$ plane depends on third order terms 
in a normal expansion of the dynamical system around the steady state. 
Therefore, this property offers a particularly stringent 
test of  constitutive equations.
We have performed the analysis of the system~(\ref{eqn:stress}),~(\ref{eqn:vf:1}) 
and~(\ref{eqn:vf:2}).
Our analysis defines a point:
\begin{equation}
\mu_{\rm crit} = {E_1\over E_0}\,{e^{-\kappa+\sqrt{2-2\kappa+\kappa^2}}\over 1-\kappa}
\left(\kappa -\sqrt{2-2\kappa+\kappa^2}\right)^2
\end{equation}
on the Hopf line, which is drawn on Figure~\ref{fig:hopf}.
Above, and on the left of this point, 
the Hopf bifurcation is supercritical (continuous);
below, and on the right of it, the bifurcation is sub-critical (discontinuous).

The shape of the phase boundary as well as the types of transitions, and even
their relative placement in the phase diagram are
consistent with current experimental observations.
However, a complete and quantitative characterization of the phase diagram for
boundary lubrication remains an open challenge  experimentally.
Overcoming obstacles associated with the limited range of stiffness which can be probed
with existing techniques
will enable key observations which can be compared with both the contrasting case of
dry friction experiments and the
theoretical results presented here.

\subsection{Chaotic stick-slip}
Drummond and Israelachvili~\cite{drummond00} observed that in some cases stick-slip
motion became erratic in the neighborhood of  the transition. These authors identified the
erratic motion as {\it chaotic}, based on the analysis of experimental time series. Their
results were  indeed suggestive of positive Lyapunov exponents. However,
methods to identify chaos  based on analysis of an individual experimental time series  are
necessarily approximate and inconclusive,
and alone are insufficient to guarantee
that the erratic motion is chaos  as opposed to some form of
noise amplification.

Given these uncertainties, models can play a special role in helping to differentiate
between mechanisms which may lead to chaos vs. other modes of irregular motion.
Here we study the full set of nonlinear constitutive 
equations~(\ref{eqn:stress}),~(\ref{eqn:stzdil:1}-\ref{eqn:stzdil:4}), which include
the STZ state variables.
The dimension of this dynamical system is three, which is the minimum 
requirement for chaos. As shown below, over a restricted range of parameters,
the model admits chaotic solutions, characterized by irregular stick-slip and
positive Lyapunov exponents.
Because these equations are deterministic, 
noise amplification is ruled out as a source of irregular motion in the model.

In order to measure Lyapunov exponents we directly integrate the equations of motion
of a Lyapunov vector $u(t)$ in the tangent space (see {\it e.g.\/}~\cite{arnold74}).
The largest Lyapunov 
exponent $\lambda$ is estimated from the long-time behavior of the norm of this vector:
\begin{equation}
\lambda = \lim_{t\to\infty}\,{1\over t}\,\log\| u(t)\|
\quad.
\end{equation}
For this numerical study, we  used a fourth order Runge-Kutta algorithm
with fixed and variable time steps. We checked that the results do not
change qualitatively as a function of the  precision of our numerical method, as long as
the precision is sufficiently high.

\begin{figure}
\begin{center}
\unitlength = 0.005\textwidth
\begin{picture}(100,110)(5,0)
\put(10,5){\resizebox{90\unitlength}{!}{\includegraphics{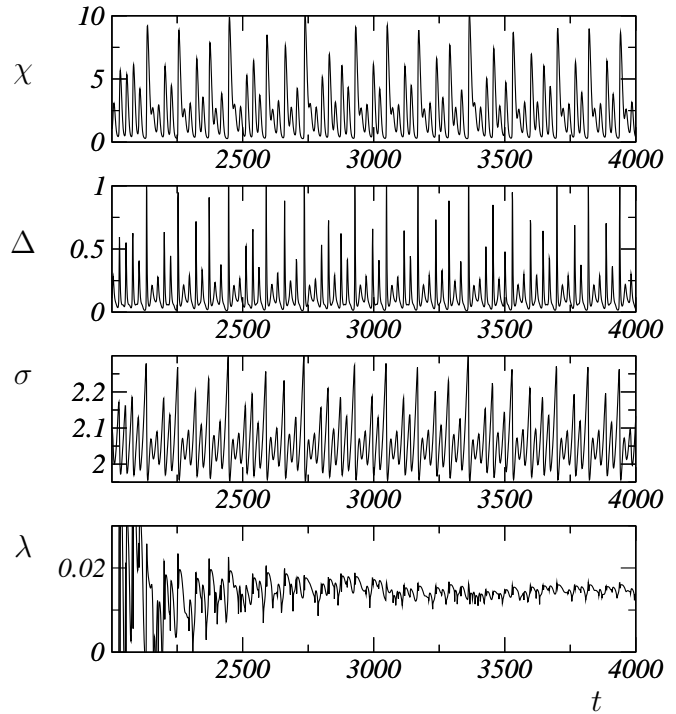}}}
\put(5,95){\makebox(0,0){\large $\chi$}}
\put(5,70){\makebox(0,0){\large $\Delta$}}
\put(5,50){\makebox(0,0){\large $\sigma$}}
\put(5,25){\makebox(0,0){\large $\lambda$}}
\put(90,2){\makebox(0,0){\large $t$}}
\end{picture}
\end{center}
\caption{ \label{fig:lyapunov}
Chaotic motion:
$\Delta$, $\chi$ , $\sigma$ and $\lambda$ are plotted as a function of time $t$
for equations~(\ref{eqn:stress}), (\ref{eqn:stzdil:1}), (\ref{eqn:stzdil:2}), (\ref{eqn:stzdil:4}).
for parameters, $E_0 = 0.5, E_1 =1, \epsilon_0=1, \alpha =2, \mu_0=0.5, 
\bar\mu=0.5,\mu=0.14, \kappa=0.8$ and $\dot\epsilon=0.14$.
The asymptotic limit reached by $\lambda(t)$ is the largest Lyapunov exponent,
which is clearly positive, indicating chaos.
}
\end{figure}

Figure~\ref{fig:lyapunov} shows typical traces in the chaotic regime. 
The three variables 
$\sigma$, $\Delta$, $\chi$, as well as the value of 
\begin{equation}
\lambda(t) \equiv {1\over t}\,\log\| u(t)\|
\quad,
\end{equation}
are shown.
This data illustrates that chaotic stick-slip 
can arise in this system. 
Chaos results from the interplay between the dynamics of the variable $\Delta$,
which characterizes the anisotropy of the molecular packing, and the variable 
$\chi$ which characterizes the dilatancy of this packing.

Further investigation reveals that the
range of parameters where chaos (i.e. at least one positive Lyapunov exponent) is
observed is restricted to a relatively compact area of the $\mu$ vs. $\dot\epsilon$
phase diagram, which is close to, but not overlapping the transition from stick-slip to steady sliding, 
in a region near the critical point separating sub- and super-critical Hopf bifurcations.
This relatively restricted range is not surprising. Since chaos requires the dynamical 
system to be three-dimensional, it may
disappear as soon as one variable is enslaved to another. In our case, this decoupling
may  occur either
when the time scale of the dynamics of $\Delta$ and $\chi$ are well separated from each other
or when the absolute value of $\Delta$ becomes negligible 
in equation~(\ref{eqn:stzdil:1}).

\begin{figure}
\begin{center}
\unitlength = 0.005\textwidth
\begin{picture}(100,70)(0,0)
\put(0,-7){\resizebox{100\unitlength}{!}{\includegraphics{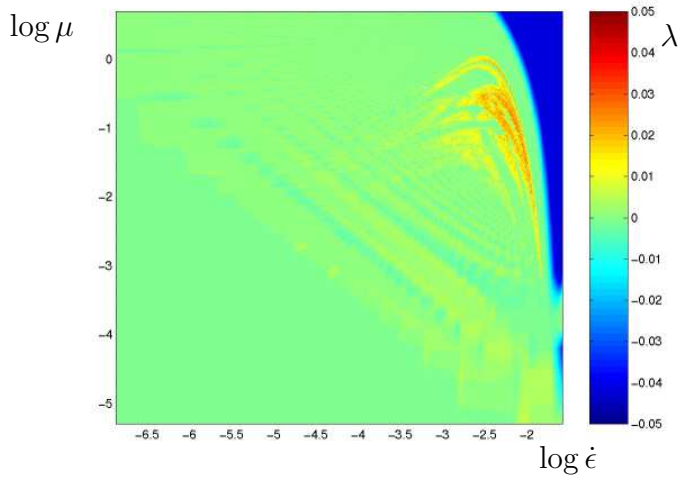}}}
\put(80,-4){\makebox(0,0){\large $\log\dot\epsilon$}}
\put(2,60){\makebox(0,0){\large $\log\mu$}}
\put(95,59){\makebox(0,0){\large $\lambda$}}
\end{picture}
\end{center}
\caption{ \label{fig:chaos}
Two-dimensional plot of the asymptotic Lyapunov exponent in the 
$(\log(\dot\epsilon), \log(\mu))$-plane obtained from numerical integration of equations
(1, 7-10). The color scale on the right gives the maximum value of 
the Lyapunov exponent $\lambda$, as defined in equation (40).
Negative values of the Lyapunov exponent appear in blue:
they correspond to steady sliding. The vertical asymptote of the phase boundary as $\log(\mu)\to -\infty$ 
(i.e. roughly the right edge of the figure) corresponds
to $\epsilon^*$. Periodic stick slip corresponds to green, corresponding to 
vanishing of the Lyapunov exponent. Chaotic behavior is associated with 
positive values of the Lyapunov exponent, which appear here in red and yellow.
Chaotic regions of stick-slip are separated by windows of periodic motion.
}
\end{figure}

To illustrate the  chaotic domain, in
Figure~\ref{fig:chaos} we present a two-dimensional image illustrating the value
of the maximum Lyapunov exponent as a function of the parameters $\mu$ and $\dot\epsilon$.
In the color map, negative values of the Lyapunov exponent appear in blue, and describe the case of a
stable fixed point in the dynamics. This
corresponds to steady sliding motion at constant $\chi$. This is the solution
between the Hopf bifurcation and the $\dot\epsilon=\dot\epsilon^*$  line
(recall for $\dot\epsilon>\epsilon^*$, $\chi$ diverges, and our
theory breaks down). 
In the color map, green corresponds to a zero value of the Lyapunov exponent.
The value $\lambda=0$  characterizes
a time translation invariant manifold.
Green thus describes stick-slip regimes, which exhibit a periodic  limit cycle.
Green also appears as the maximum Lyapunov exponent for large values of the driving 
shear rate ($\dot\epsilon>\dot\epsilon^*$).
In this case, the asymptotic state of the system
is not strictly speaking a fixed point, because $\chi$ is evolving at all times.

Chaotic motion is associated with positive Lyapunov exponents which appear as red or
yellow in Figure~\ref{fig:chaos}. 
They occur within the stick-slip portion of the 
phase diagram, near the transition between super- and sub-critical bifurcations. As shown in the figure,
in the model this behavior is 
clearly separated from the Hopf boundary by a narrow range of regular stick-slip.
The close proximity of the chaotic zone to the Hopf bifurcation may explain  the fact that
experimentally it appears that  when chaotic stick-slip
is observed, it merges continuously into the Hopf transition.
While our equations predict a regular
Hopf bifurcation to periodic stick-slip, with chaotic stick-slip resulting from secondary 
instabilities, the intermediate periodic regime occurs over a sufficiently narrow range that
it may be difficult to differentiate experimentally between this case and a case in which 
the Hopf transition takes place directly via chaotic stick-slip.

\section{V. Conclusion}
\label{sec:conclusion}

We have shown that a limited set of constitutive equations  capture at least qualitatively
a wide variety of phenomena observed  experimentally in single asperity, boundary
lubricated friction.
Most of the phenomena are captured by the dynamics 
of a single state variable identified here as  free-volume.
In steady sliding, the free-volume dynamics account for the existence of velocity-weakening or
velocity-strenghtening friction laws, involving mixed power-laws and 
logarithmic dependences. In stop-start experiments, logarithmic increases in the peak stress
with increasing hold time 
results from the relaxation of free-volume, whereas for short hold times a latency 
time emerges from the non-linearity of the transition rates.
The dynamics of free-volume also accounts for the existence of a transition 
from steady sliding to stick-slip as drive velocity is decreased, and for the  presence of both
continuous and discontinuous transitions, 
depending on the stiffness of the driving apparatus.
Finally, the complete set of equations where free-volume couples to the 
dynamics of STZs, accounts for the emergence of chaotic behavior close to 
the stick-slip transition.

Our present work is in sufficiently close qualitative agreement with experiments to warrant
a further round of more quantitative experimental vs.~theoretical comparisons. Our work
leads to clear predictions for the friction in
steady sliding regimes, and the transition to stick-slip
as a function of the compliance.
Available experimental data provides
steady sliding friction over a limited range of velocities,
and a collection of stick-slip cycles at low velocities for a given interfacial material,
most often for a unique value of the compliance. 
Available measurements of power-law, or logarithmic relationships between 
force and driving velocity are not sufficient to validate or invalidate
a set of constitutive equations.
In principle, we could attempt to directly fit cyclic stick-slip data,
but in practice, this is extremely difficult, requiring additional 
assumptions about initial values of one or more internal variables.
What information can we thus use to compare theory and experiments?
Compliance is included in the model at a low cost in terms of 
the assumptions required, yet provides a large panel of 
predictions which magnify the sensitivities of the constitutive equations and 
the microscopic assumptions of the theory.
The location of the transition to stick-slip, its dependence on compliance,
and the type of transitions to stick-slip are important predictions 
of our model that present opportunities for  more detailed  comparisons with
experimental data.
Drummond and Israelachvili~\cite{drummond00} have been able to study transition to stick-slip
for a few values of the compliance over a limited range. 
We hope this initial,  exploratory study will be followed
by a more complete set measurements to compare with our predictions.

Recent experimental results by Gourdon and Israelachvili~\cite{gourdon03}
have focused on the temperature and pressure dependence of the transition to stick-slip.
While these variables are more accessible experimentally than compliance, they
present greater challenges for theory, and 
cannot be addressed directly here  
because temperature and pressure do not appear explicitly in the constitutive laws we consider.
However,
a recent derivation of similar equations~\cite{lemaitre03} provides
new theoretical insights for 
temperature and pressure dependence, suggesting future opportunities to extend our current analysis 
in this direction.

A quantity which arose in several contexts throughout this paper was the
critical strain rate $\dot\epsilon^*$, beyond which the value of $\chi$
cannot reach steady state. In our equations, above $\dot\epsilon^*$,
$\chi$
diverges with time, because the internal (bounded transition rate)
relaxation
dynamics cannot keep pace with the rate energy is added to the system.
While an actual divergence of $\chi$ may be impeded by various physical
mechanisms (which we did not attempt to incorporate), we believe that
the existence of a change of behavior at some high strain rate is physically
meaningful, and likely to be an important 
property that it is closely related to the existence of a
latency time in stop-start tests.
Because transformation rates are bounded, at high strain rates,
the underlying state variable is expected
to decouple from the relation between stress and strain rates.
This decoupling and has several consequences:
(i) since the shear rate is only weakly dependent on state variables,
and since it is an increasing function of the stress, velocity
strenghening behavior emerges.
(ii) The state variable is hidden, its  value cannot be deduced  from
the relation
between stress and strain rate, and measurements of apparent steady relationships
between stress and strain rate
may not correspond to a true steady state.
(iii) The hidden value of the state variable depends on the strain.
(iv) When the deformation is interrupted, the state variable requires
some time before
it reaches a sufficiently low value  for it to make a measurable impact
that can be probed by the emergence of a  transient peak (i.e. a stiction spike) in stress.
In this case,
the latency time becomes a function of the overall strain.
If a high strain rate fixed point is reached by the state variable,
this dependence disappears, and the latency time should only depend on
strain rate.
It would therefore be particularly interesting to see whether the
underlying
dynamics of a state variable can be probed in experiments
close to and above $\dot\epsilon^*$ and whether it is consistent with
either
the emergence of a high strain rate fixed point,
or with unstationary dynamics of the state of the lubricated contact.
Velocity strengthening has not been observed with the SFA,
probably because of intrinsic limits due to the finite scan length.
Therefore, it is likely that the high strain rate regimes we discuss
here
cannot be directly accessed with this experimental setup.
However, velocity strengthening  has been observed in
numerics,~\cite{thompson90b}
and recently in the experiments~\cite{bureau02} by Bureau and coworkers.
This latter experimental set-up is constructed to allow  sliding over
longer distances, and provides access to higher drive velocities than
those  available
with the SFA. It thus may present opportunities to probe 
some phenomena which arise at high velocities.

Finally, we note that the constitutive equations presented here are 
derived from heuristic assumptions, 
which extend STZ theory to the rearrangements occurring inside the sheared material 
in a lubricated contact.
Although our approach is phenomenological, it relies on a specific picture
of the microscopic and mesoscopic mechanisms of deformation, and thus
forms a bridge between macroscopic, empirical, descriptions
and a fundamental understanding of the microscopic physics.
The STZ picture of plastic rearrangements occurring in localized zones is supported
by numerical observations~\cite{falk98}.
However,  it would be useful to also 
image rearrangements in experimental situations.
Although the SFA provides a very well-controlled environment
at the microscopic level, it seems unlikely that such  atomistic imaging of
the lubricant could be performed. 
Other materials are more likely candidates for  such observations. Indeed,
the picture of elementary rearrangements which emerges here
is not limited to lubricants, but is expected to apply to a wide range
of amorphous materials, in particular colloidal suspensions and granular materials.
Indeed, recent experiments by Gollub have imaged local rearrangements in granular bead
packs forming the interfacial material in friction measurements. These experiments
also exhibit stick-slip at low drive velocities, and a transition to steady sliding
as the drave rate is increased.~\cite{nasuno97,nasuno98,geminard99}
A further theoretical challenge associated with these measurements arises because 
slip does not occur homogeneously in the material, but rather
is primarily restricted to a relatively narrow dilatent region near the top of the bead pack.
We expect the results presented here may extend 
to other types of amorphous materials, but in situations when strain localization does
not preempt the application of a homogeneous description. 
Colloids and granular materials may provide simultaneous access to
macrocopic rheological properties, and microscopic imaging of elementary rearrangements.

This work was supported by the W. M. Keck Foundation,
and the NSF Grant No. DMR-9813752,
and EPRI/DoD through the Program on Interactive Complex Networks.

\appendix

\section{Appendix I: Hopf analysis}
\label{app1}
We present here the details of the calculation of the Hopf bifurcation.
For the stress-linear free-volume 
constitutive equations~(\ref{eqn:vflin:1}) and~(\ref{eqn:vflin:2}).
The complete dynamical studied can be rewritten as:
\begin{eqnarray}
\label{eqn:dyn:1}
\dot\sigma &=& \mu\,(\dot\epsilon-E_0\,\exp\left[-{1\over\chi}\right]\,\sigma)\\
\label{eqn:dyn:2}
\dot\chi &=& -E_1\,\exp\left[-{\kappa\over\chi}\right] + \alpha \,E_0\,\exp\left[-{1\over\chi}\right]\,\sigma^2
\end{eqnarray}
The calculation of the Hopf bifurcation point is straightforward: 
it amounts to considering the trace of the Jacobian of this dynamical system.
However, the calculation identifying the critical Hopf point $\mu_{crit}$ 
dividing the sub- and super-critical
bifurcation lines
requires more lengthy calculations and benefits from the introduction of  simplifying
notation. Let us write the equations as:
\begin{eqnarray}
\label{eqn:gen:1}
\dot\sigma &=& \mu\,(\dot\epsilon-f_1(\chi)\,\sigma)\\
\label{eqn:gen:2}
\dot\chi &=& -f_2(\chi) + \alpha\,f_1(\chi)\,\sigma^2
\end{eqnarray}
and perform the analysis in this more general framework.

The stationary solution is determined by
\begin{equation}
\sigma^2\equiv S(\chi) = \frac{f_2(\chi)}{\alpha f_1(\chi)}
\quad,
\end{equation}
and
\begin{equation}
\dot\epsilon^2 \equiv E(\chi) = \frac{f_1(\chi)\,f_2(\chi)}{\alpha}
\end{equation}
where the functions $E$ and $S$ have been introduced for future convenience.
In the case of equations~(\ref{eqn:dyn:1}) and~(\ref{eqn:dyn:2}),
these functions are:
\begin{equation}
E(\chi) =  {E_0 E_1\over \alpha}\exp\left[-{\kappa+1\over\chi}\right]
\quad,
\end{equation}
and
\begin{equation}
S(\chi) = {E_1\over\alpha E_0}\exp\left[-{\kappa-1\over\chi}\right]
\quad.
\end{equation}

The Jacobian of this dynamical system reads:
\begin{equation}
J = \left(
\matrix{
-\mu\,f_1(\chi) & -\mu\,f'_1(\chi)\,\sigma\cr
2\,\sigma\,f_1(\chi) & -f'_2(\chi)+\sigma^2\,f'_1(\chi)
}
\right)
\quad,
\end{equation}
and around the stationary solution, the eigenvalues of the Jacobian satisfy:
\begin{equation}
\lambda^2 +\lambda\;\sqrt{\frac{E(\chi)}{S(\chi)}}\left(\mu+\alpha\,S'(\chi)\right)
+\alpha\,\mu\,E'(\chi)=0
\quad.
\end{equation}
The Hopf bifurcation occurs when 
\begin{equation}
\mu = - \alpha S'(\chi) 
\quad,
\end{equation}
at any point where $E(\chi)$ is strictly increasing.
In our case, $E$ is always an increasing function of $\chi$, and $S$ is a decreasing 
function of $\chi$ if and only if $\kappa<1$. 
If $S$ is an increasing function of $\chi$ ($\kappa>1$) -- which also means 
that $S\circ E^{-1}$ is an increasing function of $\dot\epsilon$,
there is no Hopf bifurcation, and the steady sliding motion is stable. 
If $\kappa<1$, $S$ is a decreasing function of $\chi$;
for any $\dot\epsilon$, there is a critical value of $\mu$
where the system undergoes a Hopf bifurcation:
\begin{eqnarray*}
\mu_{\rm hopf} 
&=& -\alpha S'\left(E^{-1}(\dot\epsilon^2)\right)\\
&=& {E_1\over E_0}\,{1-\kappa\over(\kappa+1)^2}\;
\left({\alpha\,\dot\epsilon^2\over E_0\,E_1}\right)^{{\kappa-1\over\kappa+1}}\;
\ln\left[{\alpha\,\dot\epsilon^2\over E_0\,E_1}\right]^2
\quad.
\end{eqnarray*}

In order to determine the type of Hopf bifurcation (super- or sub-critical),
we write our non-linear system of ODE's in normal form. 
For this purpose, we look for the 
linear operator $T$ which transforms the Jacobian as:
\begin{equation}
T^{-1}\,J\,T=\left(\matrix{\rho & -\omega\cr \omega&\rho}\right)
\end{equation}
where the eigenvalues of the Jacobian are $\lambda_\pm =\rho\pm i\omega$. 
We also have
\begin{equation}
\rho = -\frac{1}{2}\,\sqrt{\frac{E(\chi)}{S(\chi)}}\left(\mu+\alpha S'(\chi)\right)
\end{equation}
and
\begin{equation}
\omega^2 = -\rho^2 + \alpha\mu\,E'(\chi)
\quad.
\end{equation}
The complex eigenvectors associated with $\lambda_\pm$ are 
\begin{equation}
u_\pm=
\left(\matrix{
-\frac{\rho\pm i\omega}{2\alpha\sqrt{E(\chi)}}-\frac{\mu}{2\alpha\sqrt{S(\chi)}}\cr\cr
1}
\right)
\quad,
\end{equation}
and $T$ is obtained from the real and imaginary parts of these complex eigenvectors,
$u_r={\rm Re}(u_+), u_i={\rm Im}(u_+)$
\begin{eqnarray*}
T &=& {(u_r\,u_i)}\\
&=&
\left(\matrix{
-\frac{1}{2\alpha}\left(\frac{\mu}{\sqrt{S(\chi)}}+\frac{\rho}{\sqrt{E(\chi)}}\right)&
-\frac{1}{2\alpha}\frac{\omega}{\sqrt{E(\chi)}}\cr\cr
1 & 0}\right)
\end{eqnarray*}
and 
\begin{equation}
T^{-1}=\left(\matrix{0&1\cr\cr
-2\alpha\frac{\sqrt{E(\chi)}}{\omega}&
-\frac{\rho}{\omega}-\frac{\mu}{\omega}\,\sqrt{\frac{E(\chi)}{S(\chi)}}
}\right)
\quad.
\end{equation}
The condition $\rho=0$ determines the Hopf bifurcation,
and at this point, the transformations $T$ and $T^{-1}$ reduce to,
\begin{equation}
T=\left(\matrix{
\frac{S'(\chi)}{2\sqrt{S(\chi)}}&-\frac{1}{2}\,\sqrt{-\frac{S'(\chi)E'(\chi)}{E(\chi)}}\cr
\cr
1&0
}\right)
\end{equation}
and
\begin{equation}
T^{-1}=\left(\matrix{0&1\cr\cr
2\sqrt{-\frac{E(\chi)}{S'(\chi)E'(\chi)}}&
-\sqrt{-\frac{S'(\chi)E(\chi)}{S(\chi)E'(\chi)}}
}\right)
\end{equation}
Next, we implement the linear change of variables, 
\begin{equation}
\left(\matrix{x\cr y}\right)=T^{-1}\left(\matrix{\sigma-\sigma_0\cr \chi-\chi_0}\right)
\end{equation}
which leads to the system of ODE's,
\begin{equation}
\left(\matrix{\dot x\cr \dot y}\right) = 
\left(\matrix{\rho & -\omega\cr \omega&\rho}\right)\,\left(\matrix{x\cr y}\right)
+ \left(\matrix{f(x,y)\cr g(x,y)}\right)
\end{equation}
The stability coefficient $a$ determines whether the bifurcation is super- or sub-critical.
In normal form, this coefficient can be directly obtained from the derivatives
of the functions $f$ and $g$:
\begin{eqnarray*}
a &=& \frac{1}{16}\,\left(f_{xxx}+f_{xyy}+g_{xxy}+g_{yyy}\right)\\
&+&\frac{1}{16\,\omega}
\Big(f_{xy}\left(f_{xx}+f_{yy}\right)\Big.\\
&&-g_{xy}\left(g_{xx}+g_{yy}\right)-f_{xx}g_{xx}+f_{yy}g_{yy}\Big)
\end{eqnarray*}
We then obtain, at the Hopf bifurcation point,
\begin{eqnarray*}
f_{xx} &=& \frac{\alpha\sqrt{E(\chi)}\left(S'(\chi)^2-2S(\chi)S''(\chi)\right)}{2 S(\chi)^{3/2}}\\
f_{xy} &=& \frac{\alpha\sqrt{-S'(\chi)}E'(\chi)}{2 E(\chi)}\\
f_{yy} &=& -\frac{\alpha S'(\chi)E'(\chi)}{2\sqrt{S(\chi)}\sqrt{E(\chi)}}\\
g_{xx} &=& \frac{\alpha\sqrt{-S'(\chi)}\left(E'(\chi)^2-2E(\chi)E''(\chi)\right)}
{2E(\chi)\sqrt{E'(\chi)}}\\
g_{xy} &=& -\frac{\alpha\sqrt{E(\chi)}S'(\chi)^2}{S(\chi)^{3/2}}\\
g_{yy} &=& \frac{\alpha S'(\chi)\sqrt{-S'(\chi)}\sqrt{E'(\chi)}}{2 S(\chi)}\\
f_{xxx} &=& 
-\frac{3\alpha \sqrt{E(\chi)} S'(\chi)}{4\sqrt{S(\chi)}}
\left(\frac{S'(\chi)}{S(\chi)}-\frac{E'(\chi)}{E(\chi)}\right)\\
&&\times\left(\frac{S'(\chi)}{S(\chi)}-2\frac{S''(\chi)}{S(\chi)}\right)\\
&-&\frac{\alpha\sqrt{E(\chi)}S^{(3)}(\chi)}{\sqrt{S(\chi)}}\\
f_{xyy} &=& \frac{\alpha S'(\chi)E'(\chi)\left(E(\chi) S'(\chi)-S(\chi)E'(\chi)\right)}{4 S(\chi)^{3/2}E(\chi)^{3/2}}\\
g_{yyy} &=& 0\\
g_{xxy} &=& \frac{\alpha S'(\chi)^2\left(E(\chi) S'(\chi)-S(\chi)E'(\chi)\right)}{2 S(\chi)^{5/2}\sqrt{E(\chi)}}
\quad,
\end{eqnarray*}
whence,
\begin{eqnarray*}
a &=&  \frac{\alpha\sqrt{E(\chi)}S''(\chi)}{32S(\chi)^{1/2}}
\Bigg(3\frac{S'(\chi)}{S(\chi)}-3 \frac{E'(\chi)}{E(\chi)} \Bigg.\\
&&\Bigg.+2\frac{E''(\chi)}{E'(\chi)}-2\frac{S^{(3)}(\chi)}{S''(\chi)}
\Bigg)
\end{eqnarray*}
We thus obtain a quite simple condition for the vanishing of $a$, which determines 
the points where the super- or sub-critical character of the Hopf bifurcation change:
\begin{equation}
\frac{S^{(3)}(\chi)}{S''(\chi)} = \frac{1}{2}\left(\frac{3S'(\chi)}{S(\chi)}-\frac{3E'(\chi)}{E(\chi)}+\frac{2E''(\chi)}{E'(\chi)}\right)
\end{equation}
With the specific functions $E$ and $S$ of our current interest, the parameter $a$ reads,
\begin{equation}
a = \frac{1}{16\chi^6}\exp\left[-\frac{\kappa}{\chi}\right]E_1(1-\kappa)
\left(\kappa-1-2\kappa\chi+2\chi^2\right)
\quad,
\end{equation}
which admits a single positive zero for $\kappa<1$: 
\begin{equation}
\chi_{\rm crit}= \frac{1}{2}\left(\kappa+\sqrt{2-2\kappa+\kappa^2}\right)
\end{equation}
This value defines a critical value for the parameter $\mu$
at the Hopf bifurcation,
\begin{eqnarray*}
\mu_{\rm crit}&=&\frac{E_1}{E_0}\frac{1-\kappa}{\chi_{\rm crit}^2}\exp\left[\frac{1-\kappa}{\chi_{\rm crit}}\right]\\
&=& {E_1\over E_0}\,{e^{-\kappa+\sqrt{2-2\kappa+\kappa^2}}\over 1-\kappa}
\left(\kappa -\sqrt{2-2\kappa+\kappa^2}\right)^2
\end{eqnarray*}
and the critical strain rate,
\begin{eqnarray*}
\dot\epsilon_{\rm crit}&=&\sqrt{\frac{E_1 E_0}{\alpha}}\,
\exp\left[-\frac{\kappa+1}{\kappa-1}
\left(\kappa-\sqrt{2-2\kappa+\kappa^2}\right)
\right]
\end{eqnarray*}


\end{document}